\documentclass[twocolumn]{aastex63}

\usepackage{amsmath}
\usepackage{xspace}

\usepackage[T1]{fontenc}

 \newcommand{\E}[1]{\ensuremath{\times 10^{#1}}}

\newcommand{\swift}{{\it Swift}\xspace}

\newcommand{\target}{EP240408a\xspace}
\newcommand{\fluxcgs}{ergs~s$^{-1}$~cm$^{-2}$\xspace}

\def\gtrsim{\mathrel{\hbox{\rlap{\hbox{\lower4pt\hbox{$\sim$}}}\hbox{$>$}}}}

\def \arcsec {\hbox{$^{\prime\prime}$}}

\definecolor{blazeorange}{rgb}{1.0, 0.4, 0.0}
\definecolor{seagreen}{rgb}{0.18, 0.55, 0.34}
\definecolor{rufous}{rgb}{0.66, 0.11, 0.03}
\definecolor{royalfuchsia}{rgb}{0.79, 0.17, 0.57}
\definecolor{scarlet}{rgb}{1.0, 0.13, 0.0}
\definecolor{royalpurple}{rgb}{0.47, 0.32, 0.66}

\newcommand{\nicer}{\textit{NICER}\,}

\received{\today}
\revised{--}
\accepted{--}

\submitjournal{ApJL}

\shorttitle{EP240408a: A peculiar extragalactic transient}
\shortauthors{O'Connor et al.}
\graphicspath{{./}{figures/}}

\begin{document}

\title{
Characterization of a peculiar Einstein Probe transient EP240408a: an exotic gamma-ray burst or an abnormal jetted tidal disruption event? 
}

\correspondingauthor{Brendan O'Connor}
\email{boconno2@andrew.cmu.edu}

\author[0000-0002-9700-0036]{Brendan O'Connor}
    \altaffiliation{McWilliams Fellow}
    \affiliation{McWilliams Center for Cosmology and Astrophysics, Department of Physics, Carnegie Mellon University, Pittsburgh, PA 15213, USA}
\author[0000-0003-1386-7861]{Dheeraj Pasham}
    \affiliation{MIT Kavli Institute for Astrophysics and Space Research, Cambridge, MA 02139, USA}
\author[0000-0002-8977-1498]{Igor Andreoni}
    \affiliation{Joint Space-Science Institute, University of Maryland, College Park, MD 20742, USA}
    \affiliation{Department of Astronomy, University of Maryland, College Park, MD 20742, USA}
    \affiliation{Astrophysics Science Division, NASA Goddard Space Flight Center, Mail Code 661, Greenbelt, MD 20771, USA}
    \affiliation{University of North Carolina at Chapel Hill, 120 E. Cameron Ave., Chapel Hill, NC 27514, USA}
\author[0000-0002-8548-482X]{Jeremy Hare}
    \affiliation{Astrophysics Science Division, NASA Goddard Space Flight Center, 8800 Greenbelt Rd, Greenbelt, MD 20771, USA}
    \affiliation{Center for Research and Exploration in Space Science and Technology, NASA/GSFC, Greenbelt, Maryland 20771, USA}
    \affiliation{The Catholic University of America, 620 Michigan Ave., N.E. Washington, DC 20064, USA}
\author[0000-0001-7833-1043]{Paz Beniamini}
    \affiliation{Department of Natural Sciences, The Open University of Israel, P.O Box 808, Ra'anana 4353701, Israel}
    \affiliation{Astrophysics Research Center of the Open university (ARCO), The Open University of Israel, P.O Box 808, Ra’anana 4353701, Israel}
    \affiliation{Department of Physics, The George Washington University, Washington, DC 20052, USA}
\author[0000-0002-1869-7817]{Eleonora Troja}
\affiliation{Department of Physics, University of Rome ``Tor Vergata'', via della Ricerca Scientifica 1, I-00133 Rome, Italy}

\author[0000-0003-4631-1528]{Roberto Ricci}
\affiliation{Department of Physics, University of Rome ``Tor Vergata'', via della Ricerca Scientifica 1, I-00133 Rome, Italy}
\affiliation{INAF-Istituto di Radioastronomia, Via Gobetti 101, I-40129 Bologna, Italy}

\author[0000-0003-0699-7019]{Dougal Dobie}
\affiliation{Sydney Institute for Astronomy, School of Physics, The University of Sydney, New South Wales 2006, Australia}
\affiliation{ARC Centre of Excellence for Gravitational Wave Discovery (OzGrav), Hawthorn, Victoria, Australia}

\author[0000-0002-0568-6000]{Joheen Chakraborty}
    \affiliation{MIT Kavli Institute for Astrophysics and Space Research, Cambridge, MA 02139, USA}

\author[0000-0002-0940-6563]{Mason Ng}
    \affiliation{Department of Physics, McGill University, 3600 rue University, Montréal, QC H3A 2T8, Canada}
    \affiliation{Trottier Space Institute, McGill University, 3550 rue University, Montréal, QC H3A 2A7, Canada}

\author[0000-0002-0786-7307]{Noel Klingler}
    \affiliation{Center for Space Sciences and Technology, University of Maryland, Baltimore County, Baltimore, MD 21250, USA}
    \affiliation{Astrophysics Science Division, NASA Goddard Space Flight Center, 8800 Greenbelt Rd, Greenbelt, MD 20771, USA}
    \affiliation{Center for Research and Exploration in Space Science and Technology, NASA/GSFC, Greenbelt, Maryland 20771, USA}

\author[0000-0003-2758-159X]{Viraj Karambelkar}
\affiliation{Division of Physics, Mathematics and Astronomy, California Institute of Technology, Pasadena, CA 91125, USA}

\author[0000-0003-4725-4481]{Sam Rose}
\affiliation{Division of Physics, Mathematics and Astronomy, California Institute of Technology, Pasadena, CA 91125, USA}

\author[0000-0001-6797-1889]{Steve Schulze}
\affiliation{Center for Interdisciplinary Exploration and Research in Astrophysics (CIERA), Northwestern University, 1800 Sherman Ave, Evanston, IL 60201, USA.}

\author[0000-0001-9068-7157]{Geoffrey Ryan}
\affiliation{Perimeter Institute for Theoretical Physics, Waterloo, Ontario N2L 2Y5, Canada}

\author[0000-0001-6849-1270]{Simone Dichiara}
\affiliation{Department of Astronomy and Astrophysics, The Pennsylvania State University, 525 Davey Lab, University Park, PA 16802, USA}

\author[0000-0002-4754-3526]{Itumeleng Monageng}
    \affiliation{South African Astronomical Observatory, P.O. Box 9, Observatory 7935, Cape Town, South Africa}
     \affiliation{Department of Astronomy, University of Cape Town, Private Bag X3, Rondebosch 7701, South Africa}

\author[0000-0002-7004-9956]{David Buckley}
    \affiliation{South African Astronomical Observatory, P.O. Box 9, Observatory 7935, Cape Town, South Africa}
    \affiliation{Southern African Large Telescope, P.O. Box 9, Observatory 7935, Cape Town, South Africa}
    \affiliation{Department of Astronomy, University of Cape Town, Private Bag X3, Rondebosch 7701, South Africa}
    \affiliation{Department of Physics, University of the Free State, P.O. Box 339, Bloemfonein 9300, South Africa}

\author[0000-0001-7201-1938]{Lei Hu}
\affiliation{McWilliams Center for Cosmology and Astrophysics, Department of Physics, Carnegie Mellon University, Pittsburgh, PA 15213, USA}

\author[0000-0002-6428-2700]{Gokul P. Srinivasaragavan}
\affiliation{Department of Astronomy, University of Maryland, College Park, MD 20742, USA}
\affiliation{Joint Space-Science Institute, University of Maryland, College Park, MD 20742, USA}
 \affiliation{Astrophysics Science Division, NASA Goddard Space Flight Center, 8800 Greenbelt Rd, Greenbelt, MD 20771, USA}


\author[0000-0002-5182-6289]{Gabriele Bruni}
\affiliation{INAF -- Istituto di Astrofisica e Planetologia Spaziali, via del Fosso del Cavaliere 100, Roma, 00133, Italy}

\author[0000-0002-1270-7666]{Tom\'as Cabrera}
\affiliation{McWilliams Center for Cosmology and Astrophysics, Department of Physics, Carnegie Mellon University, Pittsburgh, PA 15213, USA}

 \author[0000-0003-1673-970X]{S. Bradley Cenko}
\affiliation{Astrophysics Science Division, NASA Goddard Space Flight Center, 8800 Greenbelt Rd, Greenbelt, MD 20771, USA}
\affiliation{Joint Space-Science Institute, University of Maryland, College Park, MD 20742, USA}

\author[0000-0002-8680-8718]{Hendrik van Eerten}
\affiliation{Department of Physics, University of Bath, Claverton Down, Bath, BA2 7AY, UK}

\author[0009-0006-7990-0547]{James Freeburn}
\affiliation{Centre for Astrophysics and Supercomputing, Swinburne University of Technology, John St, Hawthorn, VIC 3122, Australia}
\affiliation{ARC Centre of Excellence for Gravitational Wave Discovery (OzGrav), John St, Hawthorn, VIC 3122, Australia}

\author[0000-0002-5698-8703]{Erica Hammerstein}
\affiliation{Department of Astronomy, University of California, Berkeley, CA 94720-3411, USA}

\author[0000-0002-5619-4938]{Mansi Kasliwal}
\affiliation{Division of Physics, Mathematics and Astronomy, California Institute of Technology, Pasadena, CA 91125, USA}

\author[0000-0003-1443-593X]{Chryssa Kouveliotou}
\affiliation{Department of Physics, The George Washington University, Washington, DC 20052, USA}

\author[0009-0000-4830-1484]{Keerthi Kunnumkai}
\affiliation{McWilliams Center for Cosmology and Astrophysics, Department of Physics, Carnegie Mellon University, Pittsburgh, PA 15213, USA}

\author[0000-0002-9415-3766]{James K. Leung}
\affiliation{David A. Dunlap Department of Astronomy and Astrophysics, University of Toronto, 50 St. George Street, Toronto, ON M5S 3H4, Canada}
\affiliation{Dunlap Institute for Astronomy and Astrophysics, University of Toronto, 50 St. George Street, Toronto, ON M5S 3H4, Canada}
\affiliation{Racah Institute of Physics, The Hebrew University of Jerusalem, Jerusalem 91904, Israel}

\author[0000-0002-7851-9756]{Amy Lien}
\affiliation{University of Tampa, Department of Physics and Astronomy, 401 W. Kennedy Blvd, Tampa, FL 33606, USA}

\author[0000-0002-6011-0530]{Antonella Palmese}
\affiliation{McWilliams Center for Cosmology and Astrophysics, Department of Physics, Carnegie Mellon University, Pittsburgh, PA 15213, USA}

\author[0000-0001-6276-6616]{Takanori Sakamoto}
\affiliation{Department of Physics and Mathematics, Aoyama Gakuin University, 5-10-1 Fuchinobe, Chuo-ku, Sagamihara-shi Kanagawa 252-5258, Japan}




\begin{abstract}
We present the results of our multi-wavelength (X-ray to  radio) follow-up campaign of the \textit{Einstein Probe} transient EP240408a. The initial 10 s trigger displayed bright soft X-ray ($0.5$\,$-$\,$4$ keV) radiation with peak luminosity $L_\textrm{X}$\,$\gtrsim$\,$10^{49}$ ($10^{50}$) erg s$^{-1}$ for an assumed redshift $z$\,$\gtrsim$\,$0.5$ ($2.0$). The \textit{Neil Gehrels Swift Observatory} and \textit{Neutron star Interior Composition ExploreR} discovered a fading X-ray counterpart lasting for $\sim$5 d (observer frame), which showed a long-lived ($\sim$4 d) plateau-like emission ($t^{-0.5}$) before a sharp powerlaw decline ($t^{-7}$). The plateau emission was in excess of $L_\textrm{X}$\,$\gtrsim$\,$10^{46}$ ($10^{47}$) erg s$^{-1}$ at $z$\,$\gtrsim$\,$0.5$ ($2.0$). Deep optical and radio observations resulted in non-detections of the transient. Our observations with Gemini South revealed a faint potential host galaxy ($r$\,$\approx$\,$24$ AB mag) near the edge of the X-ray localization. The faint candidate host, and lack of other potential hosts ($r$\,$\gtrsim$\,$26$ AB mag; $J$\,$\gtrsim$\,$23$ AB mag), implies a higher redshift origin ($z$\,$\gtrsim$\,$0.5$), which produces extreme X-ray properties that are inconsistent with many known extragalactic transient classes. In particular, the lack of a bright gamma-ray counterpart, with the isotropic-equivalent energy ($10$\,$-$\,$10,000$ keV) constrained by \textit{GECam} and \textit{Konus-Wind} to $E_{\gamma,\textrm{iso}}$\,$\lesssim$\,$4\times10^{50}$ ($6\times10^{51}$) erg at $z$\,$\approx$\,$0.5$ ($2.0$), conflicts with known gamma-ray bursts (GRBs) of similar X-ray luminosities. We therefore favor a jetted tidal disruption event (TDE) as the progenitor of EP240408a at $z$\,$\gtrsim$\,$1.0$, possibly caused by the disruption of a white dwarf by an intermediate mass black hole. The alternative is that EP240408a may represent a new, previously unknown class of transient. 
\end{abstract}



\section{Introduction}
\label{sec: intro}

The extragalactic high-energy transient sky, spanning from X-rays to gamma rays, is a diverse collection of phenomena with timescales typically ranging from seconds to days. Among the most common are gamma-ray bursts (GRBs; \citealt{Kouveliotou1993}), which arise either from the merger of two compact objects \citep{Abbott+17-GW170817A-MMO,Savchenko2017,Goldstein2017} or the collapse of massive stars \citep{Woosley1993,Macfadyen1999}. GRBs exhibit two distinct phases: an initial prompt gamma-ray emission lasting typically on the order of seconds to tens of seconds \citep{Kouveliotou1993}, and an afterglow produced by the newly launched jet and its interaction with the surrounding medium \citep{Meszaros1997,Wijers1999}.

Dedicated space telescopes such as {\it Fermi} \citep{Meegan2009} and {\it Swift} \citep{Gehrels2004} have revolutionized the field. With hundreds of high-energy transients detected by these missions, some have displayed unusual behaviors. For instance, many GRBs exhibit an extended plateau phase following the prompt emission, lasting from a few hundred seconds to several hours before rapidly declining in X-rays \citep[e.g.,][]{Zhang2006,Troja2007,Rowlinson2010}. This plateau phase is thought to result from long-lived central engine activity, potentially due to the spin-down of a newborn magnetar's dipole field \citep{Zhang2006,Liang2006,Troja2007,Lyons2010,Metzger2011,Rowlinson2010,Rowlinson2013}, though other interpretations exist \citep{Shen2012,Duffell2015,Beniamini2017,BeniaminiGiannios2017,Beniamini2019plateau,Oganesyan2020,Dereli-Begue2022}. 

In 2011, \textit{Swift} identified three peculiar gamma-ray transients, later found to have long-lasting X-ray emission, with peak X-ray luminosities of $10^{47-49}$ erg s$^{-1}$ at a few days after discovery, several orders of magnitude higher than typical GRB afterglows at a similar time \citep{Bloom2011,Levan2011,Burrows2011,Zauderer2011,Burrows2011,Cenko2012,Brown2015,Pasham2015}. These events also featured prolonged high-energy activity, lasting months instead of hours -- far exceeding the duration of the longest GRB plateau phases \citep[e.g.,][]{Levan2014,Cucchiara2015}. These systems have been hypothesized to be relativistic jetted Tidal Disruption Events (TDEs) where a star is disrupted by a massive black hole and the resulting jet is fortuitously pointed along our line of sight akin to a Blazar \citep{Bloom2011,Levan2011,Cenko2012,Andreoni2022,Pasham2023,Rhodes2023,Yao2024}.

The launch of the \textit{Einstein Probe} \citep{EP2015,EP2022} has opened a new window into the soft X-ray Universe, especially for fast transients. The Wide-field X-ray Telescope (WXT), with a field of view of approximately 3,600 square degrees, monitors soft X-rays in the $0.5$\,$-$\,$4$ keV range. Since February 2024, EP has detected a variety of intriguing fast X-ray transients \citep{Zhang2024epatel}. While a number of fast transients discovered by EP have turned out to be Galactic (e.g., stellar flares or X-ray binaries) in nature \citep[e.g.,][]{Potter2024,Gaudin2024}, there are a variety that show clear GRB-like properties (e.g., EP240315a/GRB 240315C, \citealt{Levan2024,Liu2024,Gillanders2024,Ricci2024}; EP240414a, \citealt{Srivastav2024,vanDalen2024,Bright2024,Sun2024}) or have been associated with GRBs in a post-trigger, ground-based analysis \citep[e.g.,][]{Yin2024,Liu2024}. 

In this paper, we present a multi-wavelength (ultraviolet, optical, near-infrared, radio, X-ray and gamma-ray) study of the EP discovered X-ray transient \target. \target has properties distinct from any previously known high-energy transient. We argue that it is likely extragalactic in nature and has properties that are challenging to explain either as a GRB or jetted TDE.

The manuscript is laid out as follows. 
In \S \ref{sec:data} we highlight the results of the X-ray, ultraviolet, optical, near-infrared, and radio data obtained for EP240408a. 
We analyze the multiple possible interpretations for EP240408a in \S \ref{sec: discuss} and present our conclusions in \S \ref{sec: conclusions}. 

All upper limits are presented at the $3\sigma$ level and all magnitudes are in the AB photometric system. We use a flat cosmology \citep{Planck2018}. We further adopt the standard convention $F_\nu$\,$\propto$\,$\nu^{-\beta}t^{-\alpha}$.

\section{Data Reduction and Analysis}\label{sec:data}


\subsection{Einstein Probe Trigger and Gamma-ray Constraints}

The Wide-field X-ray Telescope (WXT) onboard the \textit{Einstein Probe} (EP; \citealt{EP2015,EP2022}) discovered \target at 2024-04-08 at 17:56:30 UT or MJD 60408.747\footnote{All times in this work are relative to this trigger time.} \citep{GCN36053eptrigger}. The peak flux and  average flux in the soft X-ray ($0.5$\,$-$\,$4.0$ keV) band over the $\sim$10 s duration were reported to be $\sim$\,$1.4\times10^{-8}$ \fluxcgs and $(4.0\pm1.3)\times10^{-9}$ erg cm$^{-2}$ s$^{-1}$, respectively \citep{GCN36053eptrigger}. This roughly translates to a soft X-ray fluence of $\sim$\,$4\times10^{-8}$ erg cm$^{-2}$. 

EP240408a was not detected or localized by any other high energy monitor. {\it Konus-Wind} \citep{Aptekar1995} and {\it GECam} \citep{Li2023gecam} were observing the location of \target at the time of the EP trigger, but did not detect any emission (Dmitry Svinkin, private communication). The \textit{Fermi} Gamma-ray Burst Monitor (GBM; \citealt{Meegan2009})  and \textit{AstroSat}/CZTI \citep{Bhalerao2017} have no data covering the location of EP240408a at the EP trigger time, while EP240408a was out of the field of view of \textit{Swift}'s Burst Alert Telescope \citep[BAT;][]{Barthelmy2005} (Jimmy DeLaunay/\swift and Gaurav Waratkar/\textit{Astrosat}, private communication).

\textit{Konus-Wind} was observing the position of EP240408a for 1000 s before and after the trigger time. The 90\% confidence upper limit to the peak flux ($20$\,$-$\,$1500$ keV) is $<$\,$1.5\times10^{-7}$ erg cm$^{-2}$ s$^{-1}$ for a typical long GRB spectrum and assuming a timescale of 2.944 s (Dmitry Svinkin, private communication). 
This is consistent with the typical fluence upper limits for \textit{Konus-Wind} based on all-sky searches for GRBs associated to gravitational waves (GWs) which has typical limits of $<$\,$6\times10^{-7}$ erg cm$^{-2}$ and $<$\,$2\times10^{-7}$ erg cm$^{-2}$ for short and long GRBS, respectively, in the $20$\,$-$\,$1500$ keV energy range \citep{Ridnaia2020}.  \textit{GECam-B} reported a slightly less sensitive $3\sigma$ upper limit of $<$\,$2.4\times10^{-6}$ erg cm$^{-2}$ for the gamma-ray fluence in the $15$\,$-$\,$300$ keV energy range \citep{GCN36058gecam}. 

Despite being out of the field of view of \textit{Swift}/BAT, and therefore lacking event data, the \texttt{NITRATES} pipeline \citep{NITRATES} can constrain the gamma-ray properties. The upper limit to the peak flux is $<$\,$3.4\times10^{-7}$ erg cm$^{-2}$ s$^{-1}$ for a 3.2 s timescale in the $15$\,$-$\,$350$ keV energy band, corresponding to $<$\,$5.5\times10^{-7}$ erg cm$^{-2}$ s$^{-1}$ over $20$\,$-$\,$1500$ keV for comparison to \textit{Konus-Wind} (Jimmy DeLaunay, private communication). 

We further ran the \textit{Swift}/BAT Hard X-ray Transient Monitor \citep{Krimm2013} source detection ($15$\,$-$\,$50$ keV) and lightcurve production algorithm for a period corresponding to approximately 1 month before to 2 months after the EP trigger (MJD $60374$\,$-$\,$60461$) and found no significant detections in the daily binned lightcurve. The source was out of the BAT field of view from approximately $-19$ d to $-4$ d prior to the EP trigger.

\subsection{\textit{Swift}/XRT}\label{sec: XRT}

The position of EP240408a was observed by the \textit{Neil Gehrels Swift Observatory} (hereafter \textit{Swift}; \citealt{Gehrels2004}) X-ray Telescope \citep[XRT;][]{Burrows2005} starting on 2024-04-10 (1.4 days after the EP trigger; \citep{GCN36057xrt1}). In total observations were obtained across 5 epochs between 2024-04-10 and 2024-04-26 in Photon Counting (PC) mode for a total of $8.9$ ks, see Table \ref{tab: observationsXray}. An X-ray source is detected only in the first observation at 1.4 d with a total exposure of 1.8 ks in PC mode. The following XRT observations at 10.4, 12.4, 15.4, and 17.5 d (lasting for between $1$\,$-$\,$2.5$ ks) after the EP trigger did not result in a detection of the X-ray source. We used the {\it Swift}/XRT data products generator\footnote{\url{https://www.swift.ac.uk/user_objects/}} and upper limit server\footnote{\url{https://www.swift.ac.uk/LSXPS/ulserv.php}} \citep{LSXPS} to produce $3\sigma$ upper limits on the unabsorbed X-ray flux. 

\swift/XRT data taken 1.4 d after the EP trigger yielded a standard position \citep{Evans2009} of RA, DEC (J2000) = $10^{h}35^m 24^{s}.28$, $-35^\circ 44\arcmin 49.9\arcsec$ with an uncertainty of $3.5\arcsec$ (90\% confidence level; CL). However, a more refined enhanced position \citep{Goad2007,Evans2009} is derived as RA, DEC (J2000) = $10^{h}35^m 23^{s}.96$, $-35^\circ 44\arcmin 55.1\arcsec$ with an uncertainty of $2.2\arcsec$ (90\% CL). The enhanced position shifts by $\sim$\,$6.5\arcsec$ with respect to the initially reported standard position (Wenda Zhang, private communication). 

\subsection{\textit{NICER}}\label{sec: nicer}
EP240408a was monitored by the \textit{Neutron star Interior Composition ExploreR} (\textit{NICER}) using the X-ray timing instrument \citep[XTI;][]{Gendreau2016} between 2024-04-10 (MJD 60410.564) and 2024-05-16 (MJD 60447.259), corresponding to 1.8 to 38 d after trigger, with a near daily cadence (ObsIDs 7204340101 to 7204340131; Table \ref{tab: observationsXray}). During this period \textit{NICER} made several visits per day that typically lasted for a few hundred seconds. This resulted in an on-source exposure time of 109 ks. The initial four days of data were first reported in \cite{ATel16589saltnicer}, and have been re-analyzed here along with the full dataset. 

We retrieved the data from the \textit{NICER} data archive. The data were processed using \texttt{NICERDAS v12} within   \texttt{HEASOFT v6.34} and the latest calibration files. After retrieving the latest geomagnetic data, we applied processed the \texttt{nicerl2} task to generate cleaned event files with the default screening criteria. The late time data was severely impacted by increased solar activity and all data was impacted by a significant Oxygen line from Earth's atmosphere. We attribute this to significant background fluctuations due to solar activity and the atmospheric Oxygen line, which we discuss further in \S \ref{sec:xrayspectra}. 

\textit{NICER} data consisted of both the International Space Station's (ISS's) night and the day time data. As per the \textit{NICER} data analysis guide\footnote{\url{https://heasarc.gsfc.nasa.gov/docs/nicer/analysis_threads/}}, for each ObsID, we extracted separate ``clean'' event files for both night and day data. Then using the \texttt{nicerl3-spect} tool and the \texttt{SCORPEON} background modeling framework we extracted separate night and day spectra from individual ObsIDs. 
We discuss our spectral modeling of the \textit{NICER} data in \S \ref{sec:xrayspectra}.

\subsection{\textit{NuSTAR}}\label{sec: nustar}
We observed EP240408a through a Director's Discretionary Time request (ObsID: 91001622; PI: O'Connor) for a total of 42 ks starting on  2024-04-22 at 00:36:09 UT ($T_0$\,$+$\,$13.3$ d). The data were reduced using the \textit{NuSTAR} Data Analysis Software pipeline (\texttt{NuSTARDAS}) within \texttt{HEASOFT v6.33.2}. 
At the location of EP240408a we do not detect a source in either FPMA or FPMB. As there are no sources detected in the image we cannot correct for the pointing accuracy of \textit{NuSTAR} or any offsets between FPMA and FPMB. We extract the total observed counts using a $30\arcsec$ circular region centered on EP240408a and a nearby background region of $80\arcsec$ for both FPMA and FPMB. 
Using the formalism presented by \citet{Kraft1991}, we derive a $3\sigma$ upper limit of $<$\,$1.4\times10^{-3}$ cts s$^{-1}$ ( $3$\,$-$\,$79$ keV). Adopting an absorbed powerlaw with photon index $\Gamma$\,$=$\,$2$ and Hydrogen column density $N_H$\,$=$\,$1\times10^{21}$ cm$^{-2}$, we derive a $3\sigma$ upper limit of $<$\,$9.98\times10^{-14}$ erg cm$^{-2}$ s$^{-1}$ ( $3$\,$-$\,$79$ keV) to the unabsorbed flux of EP240408a at $\sim$13.3 d after the EP trigger.

\subsection{Archival X-ray Data}
We utilized the High-Energy Lightcurve Generator (HILIGT; \citealt{Konig2022,Saxton2022})\footnote{\url{http://xmmuls.esac.esa.int/upperlimitserver/}} to determine whether the source has experienced any archival outburst potentially caught by \textit{ROSAT}, \textit{INTEGRAL}, or the \textit{XMM-Newton} Slew Survey \citep[e.g.,][]{Saxton2008}. The source position was observed by \textit{XMM-Newton} during slews 6 times between 2005-05-20 and 2024-05-18 with short $\sim$\,$5$\,$-$\,$8$ s exposures. Each observation has a similar limit in the three energy ranges ($0.2$\,$-$\,$2$, $2$\,$-$\,$12$, and $0.2$\,$-$\,$12$ keV), but the latest observation serendipitously occurred on 2024-05-18 at 21:34:05 UT yielding a $3\sigma$ upper limit of $<$\,$2\times10^{-12}$ erg cm$^{-2}$ s$^{-1}$ in the $0.2$\,$-$\,$12$ keV energy range.
An absorbed powerlaw with photon index $\Gamma$\,$=$\,$2$ and Hydrogen column density $N_H$\,$=$\,$1\times10^{20}$ cm$^{-2}$ was used. 
The HILIGT also identified upper limits ($3\sigma$) from past \textit{ROSAT} and \textit{INTEGRAL} observations, assuming the same spectral shape, with $<$\,$2\times10^{-13}$ erg cm$^{-2}$ s$^{-1}$ ($0.2$\,$-$\,$2$ keV) from \textit{ROSAT} on 1990-10-28 and $<$\,$7\times10^{-12}$ erg cm$^{-2}$ s$^{-1}$ ($20$\,$-$\,$40$ keV) from \textit{INTEGRAL} on 2006-06-18. 
Therefore we have found no evidence for past X-ray outbursts from this source. 

We likewise checked the \textit{eROSITA} upper limit server\footnote{\url{https://erosita.mpe.mpg.de/dr1/erodat/upperlimit/single/}} \citep{erositaULs}. No source is detected at the location of the XRT source with $3\sigma$ upper limits $<$\,$6.7\times10^{-14}$ erg cm$^{-2}$ s$^{-1}$ ($0.2$\,$-$\,$6$ keV) and $<$\,$6.5\times10^{-14}$ erg cm$^{-2}$ s$^{-1}$ ($0.2$\,$-$\,$2.3$ keV). The non-detection in \textit{eROSITA} was first reported by \citep{ATel16584fxt}.

\begin{figure*}
    \centering
\includegraphics[width=1\columnwidth]{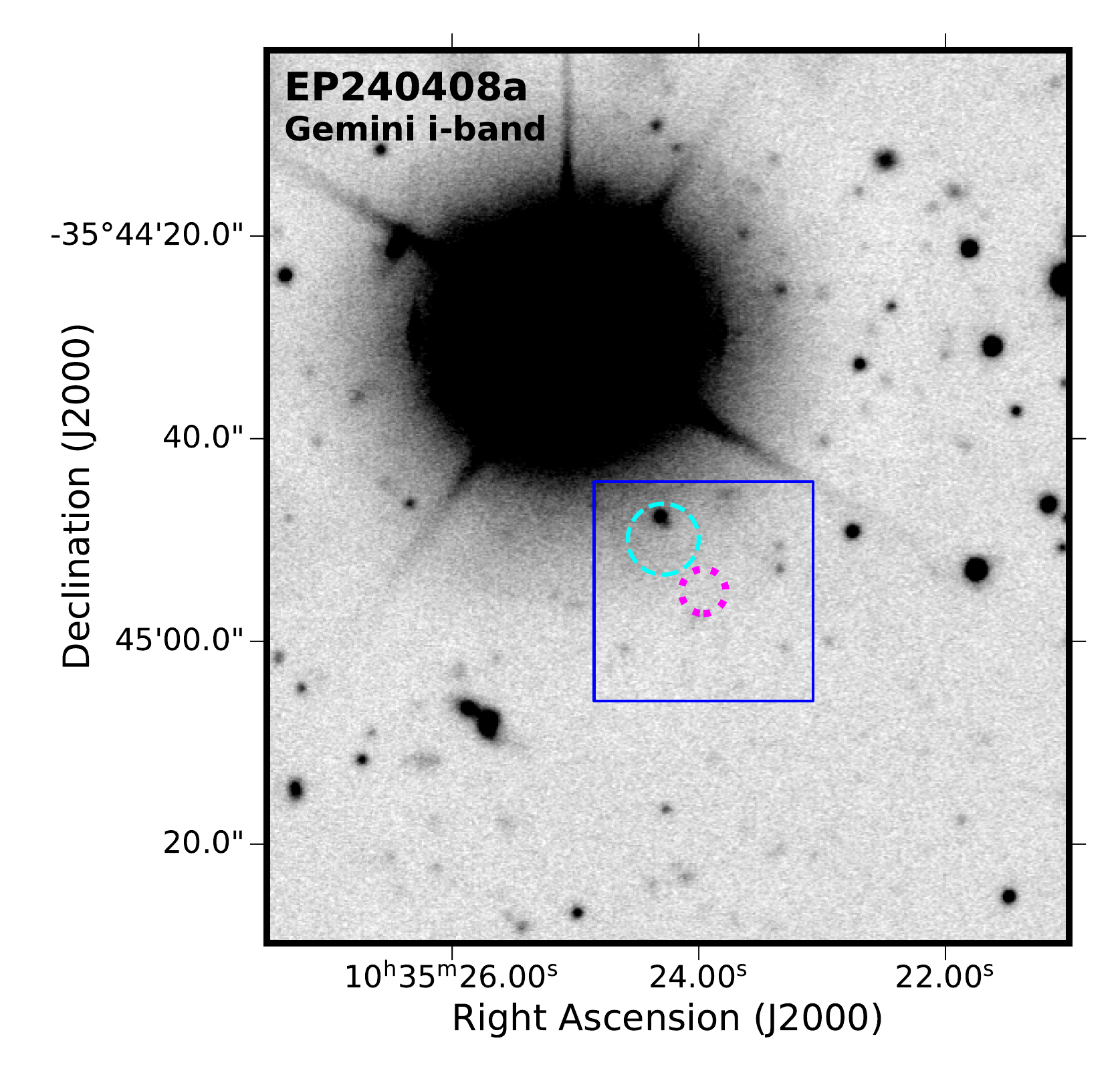}
\includegraphics[width=1\columnwidth]{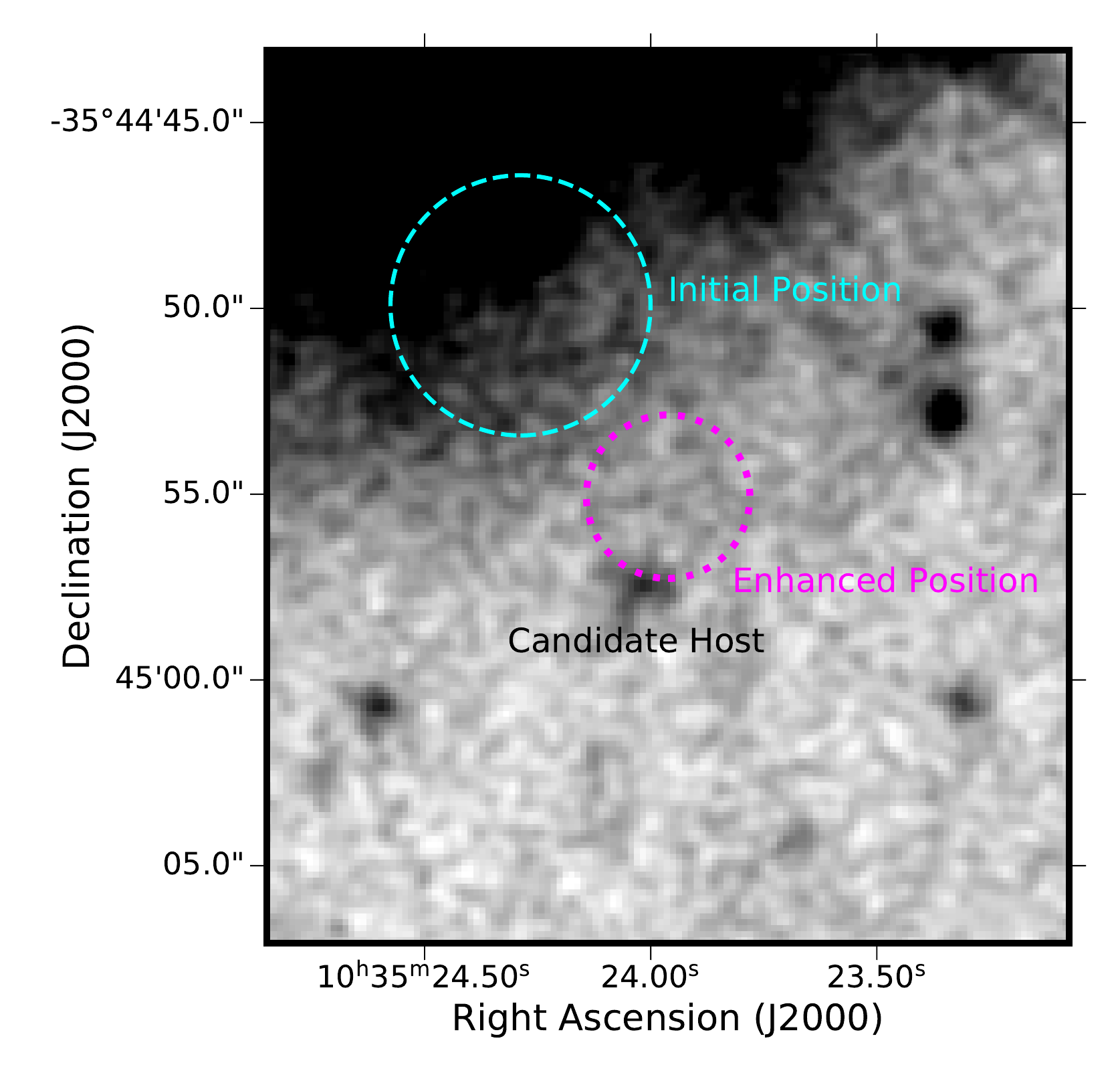}
    \caption{
    \textbf{Left:} Finding Chart of EP240408a using deep images obtained with Gemini GMOS-S in $i$-band at $T_0$\,$+$\,$82$ d (2024-06-30) with $3\sigma$ limiting depth $\gtrsim$26 AB mag. The initial XRT localization (radius $3.5\arcsec$; 90\% CL) is shown as a cyan  dashed circle. The enhanced XRT localization (radius $2.2\arcsec$; 90\% CL) is represented by a magenta dotted circle.  
    The field of view of the zoom in region (right panel) is shown by a blue square. \textbf{Right:} Zoom in on the enhanced localization of EP240408a. To the South-East of the enhanced position (dotted magenta circle) lies a candidate host galaxy. The image is smoothed for display purposes. North is up and East is to the left. } 
    \label{fig:fc}
\end{figure*}

\subsection{Gemini}\label{sec: gemini}

We observed EP240408a with Gemini-South through a DDT request (GS-2024A-DD-104; PI: O'Connor) using the Gemini Multi-Object Spectrograph (GMOS) and FLAMINGOS-2 (F2). We carried out observations across three epochs on 2024-04-25, 2024-05-02, and 2024-06-30. We observed in the $riJ$ filters with a total exposure of 720, 720, and 900 s for the first two epochs. The third epoch was perform only in $ri$ filters for 960 and 1,000 s, respectively. The third epoch serves as a template for image subtraction, and achieved the best observing conditions and seeing.

The data were reduced using standard tasks within the \texttt{Dragons} software package \citep{Labrie2019,Labrie2023}, including using the \texttt{ultradeep} recipe for F2 images. At the South-East edge of the XRT enhanced position we identify a candidate host galaxy (Figure \ref{fig:fc}). In order to gauge variability between epochs, we performed difference imaging with the Saccadic Fast Fourier Transform (\texttt{SFFT}) software\footnote{\url{https://github.com/thomasvrussell/sfft}} \citep{Hu2022}. We performed image subtraction between all epochs for each filter and identify no optical or near-infrared variability in either the initial (standard) XRT position, the enhanced XRT position, or at the location of the candidate host galaxy. Our $3\sigma$ upper limits based on using our third epoch (2024-06-30) as the template for the $ri$ filters and second epoch (2024-05-02) for $J$-band are reported in Table \ref{tab: observationsOpt}.Aperture photometry was performed using \texttt{SExtractor} \citep{Bertin1996} and photometric zeropoints calibrated to the SkyMapper \citep{Keller2007,Wolf2018} and 2MASS \cite{Skrutskie2006} catalogs.

To determine its nature and distance scale, we obtained spectroscopic observations of the bright source \citep{GCN36059grond}, see Appendix \ref{sec:counter}, lying within the initial (standard) XRT localization (Figure \ref{fig:fc}) using Gemini GMOS-S through program GS-2024A-FT-113 (PI: Andreoni) starting on 2024-06-30 at 23:59:20 UT. The data were acquired with the R400 grating with $2\times1,000$ s exposure at central wavelength $\sim8,000$ \AA\, and 
$2\times1,000$ s centered at $\sim8,200$ \AA. A $1\arcsec$ slit width was used at a position angle of $127^\circ$ chosen to minimize contamination from the nearby bright star. The data were reduced using \texttt{Dragons}. An inspection of the stacked 2D spectrum reveals no clear narrow emission lines in the observed wavelength range between $\sim$\,$5,700$\,$-$\,$10,500$ \AA (Appendix \ref{sec:counter}). The red end of the spectrum is severely impacted by bright sky emission lines.

\subsection{Keck}
A near-infrared spectrum of the bright source \citep{GCN36059grond}, see Appendix \ref{sec:counter}, within the initial standard XRT position was obtained with the 
Near-Infrared Echellette Spectrometer \citep[NIRES;][]{Wilson2004} on the Keck II telescope starting on 2024-04-19 at 06:57:36 UT ($10.54$ d post-trigger). The spectrum  ($1.0$\,$-$\,$2.4$ $\mu$m) was obtained at airmass 1.77 using the 0.55 arcsec slit with 4$\times$300 s exposures. 
Data were reduced using standard reduction procedures within \texttt{pypeit} \citep{Prochaska2020}. The spectrum has a very low signal-to-noise ratio and is dominated by sky emission lines. We heavily re-binned the spectrum (9 pixels) and find no discernible features.

\subsection{\textit{Swift}/UVOT}
As described in \S \ref{sec: XRT}, the \textit{Swift}  Ultra-Violet Optical Telescope (UVOT; \citealt{Roming2005}) observed EP240408a during the same time intervals as \textit{Swift}/XRT due to its simultaneous instrument operation.  
The exposure time was split between the various optical/UV filters: $v$, $b$, $u$, $uvw1$, $uvm2$, and $uvw2$ (Table \ref{tab: observationsOpt}). We used a circular source extraction region centered on the XRT position with a radius of $3.5''$, which is the XRT positional uncertainty and a typical source extraction radius for faint sources. 
EP240408a was not detected in any observations, in any filters. The upper limits are given in Table \ref{tab: observationsOpt}. For the background regions, we use a circular region of at least $20''$ radius, placed near EP240408a's position. The exact position and size of the background region varies between observations, as it has to be placed differently in different/stacked observations in order to order to avoid image artifacts (e.g., readout streaks, smoke rings) caused by nearby bright sources which are roll-angle and pointing dependent, as well as bright sources and their large PSFs which are present in some filters and not others. In order to obtain deeper limits, we also stacked all observations (of the same filter) from different epochs using the \texttt{HEASOFT} tool {\tt uvotimsum}. No source was detected, and these limits are given in Table \ref{tab: observationsOpt}. 

\subsection{Dark Energy Camera (DECam)}\label{sec: archival}
We retrieved publicly available imaging obtained with the Dark Energy Camera (DECam). The data were obtained through the NOIRLab Astro Data Archive\footnote{\url{https://astroarchive.noirlab.edu/}} \citep{2021Mirro...2...33M}. There is available imaging of the field of EP240408a in the $grz$ filters (see Table \ref{tab: observationsOpt} for details). Aperture photometry was performed using the \texttt{SExtractor} software and photometric zeropoints calibrated to the SkyMapper catalog. We find no source at the location of the updated (enhanced) XRT localization or at the location of our candidate host galaxy (Figure \ref{fig:fc}).   

\subsection{Very Large Array (VLA)}\label{sec: vla}
We carried out observations with the Karl J. Jansky Very Large Array (VLA) on 2024-04-19, 2024-09-13, and 2024-12-12 (24A-320; PI: O'Connor), see Table \ref{tab: observationsradio}. The observations we performed in X-band with a central frequency of 10 GHz and a bandwith of 4 GHz with the array in C configuration during the first observation (2024-04-19), B configuration on 2024-09-13, and A configuration on 2024-12-12. The time on source was $\sim$24 minutes in both observations.  
The data were retrieved from the National Radio Astronomical Observatory (NRAO) archive, and processed using the VLA \texttt{CASA} pipeline in \texttt{CASA} v6.5.4 \citep{CASA2007}. 
We used the sources 3C286 and J1051-3138 as primary and phase calibrators. \target does not show any radio emission (Table \ref{tab: observationsradio}) with deep $3\sigma$ upper limits obtained with the Very Large Array (VLA) on 2024-04-19 (11 days), 2024-09-13 (158 days), and 2024-12-12 (258 days) of $<$\,$17\,\mu$Jy, $<$\,$21\,\mu$Jy, and $<$\,$20\,\mu$Jy, respectively.

\subsection{Australia Telescope Compact Array (ATCA)}\label{sec: ATCA}
We observed the location of EP240408a with the ATCA between 2024-05-01T04:51:05 and 2024-05-01T10:32:05 (C3589; PI: Dobie). The observation was split into four groups of 2$\times$12 minute observations across that period, for 96 minutes on-source in total. Observations were carried out with $2\times2048$\,MHz bands centered on 5.5 and 9\,GHz. We used an observation of the ATCA primary calibrator, 1934$-$638, as the bandpass and flux scale calibrator and 1034$-$374 as the gain calibrator. We carried out standard ATCA data reduction with {\sc miriad} \citep{Sault1995_MIRIAD} and found no emission at the transient location with a $3\sigma$ upper limit of $60\,\mu$Jy in both bands.

EP240408a was also observed with the ATCA by an independent group under a target-of-opportunity program (CX570; PI: Shu) on two occasions. The first observation was conducted on 2024-05-08 with $2\times2048$\,MHz bands centered on 5.5 and 9\,GHz. The second observation was conducted on 2024-05-10 with $2\times2048$\,MHz bands centered on 17 and 19\,GHz. Neither observation has an associated observation of the ATCA primary calibrator (1934$-$638), nor were there any observations of it within a day of either observation. Hence, standard flux calibration of these observations is not possible. We have not analysed the 2024-05-08 data because of this, and the proximity to our more sensitive observation on 2024-05-01.

Nevertheless, we reduced the data obtained on 2024-05-10 because it provides coverage at higher observing frequencies. We used 1034$-$293 as the flux and bandpass calibrator and 1048$-$313 as the gain calibrator. Both bands were reduced independently in {\sc miriad} and then scaled to have a consistent flux scale using {\sc mfboot} before being combined and imaged as a single contiguous band. We do not detect any radio emission at the position of the transient. We set the absolute flux scale by setting the flux density of 1034$-$293 to 2\,Jy based on its typical flux density in the ATCA calibrator database\footnote{\url{https://www.narrabri.atnf.csiro.au/calibrators/calibrator_database_viewcal?source=1034-293}}. However, 1034$-$293  appears to be significantly variable (with previously observed flux densities ranging from 1-3\,Jy) and hence this scaling is likely unreliable. Based on this calibration the image noise is $\sim 15\,\mu$Jy, which is four times more sensitive than the noise estimate reported by the ATCA sensitivity calculator. We consider the ATCA sensitivity to be a more reliable (albeit, optimistic) estimate of the overall flux scale, and hence we report a $3\sigma$ upper-limit of $180\,\mu$Jy on the source flux density for this observation.

\subsection{X-ray Spectroscopy}\label{sec:xrayspectra}

\begin{figure*}
    \centering
\includegraphics[width=1.5\columnwidth]{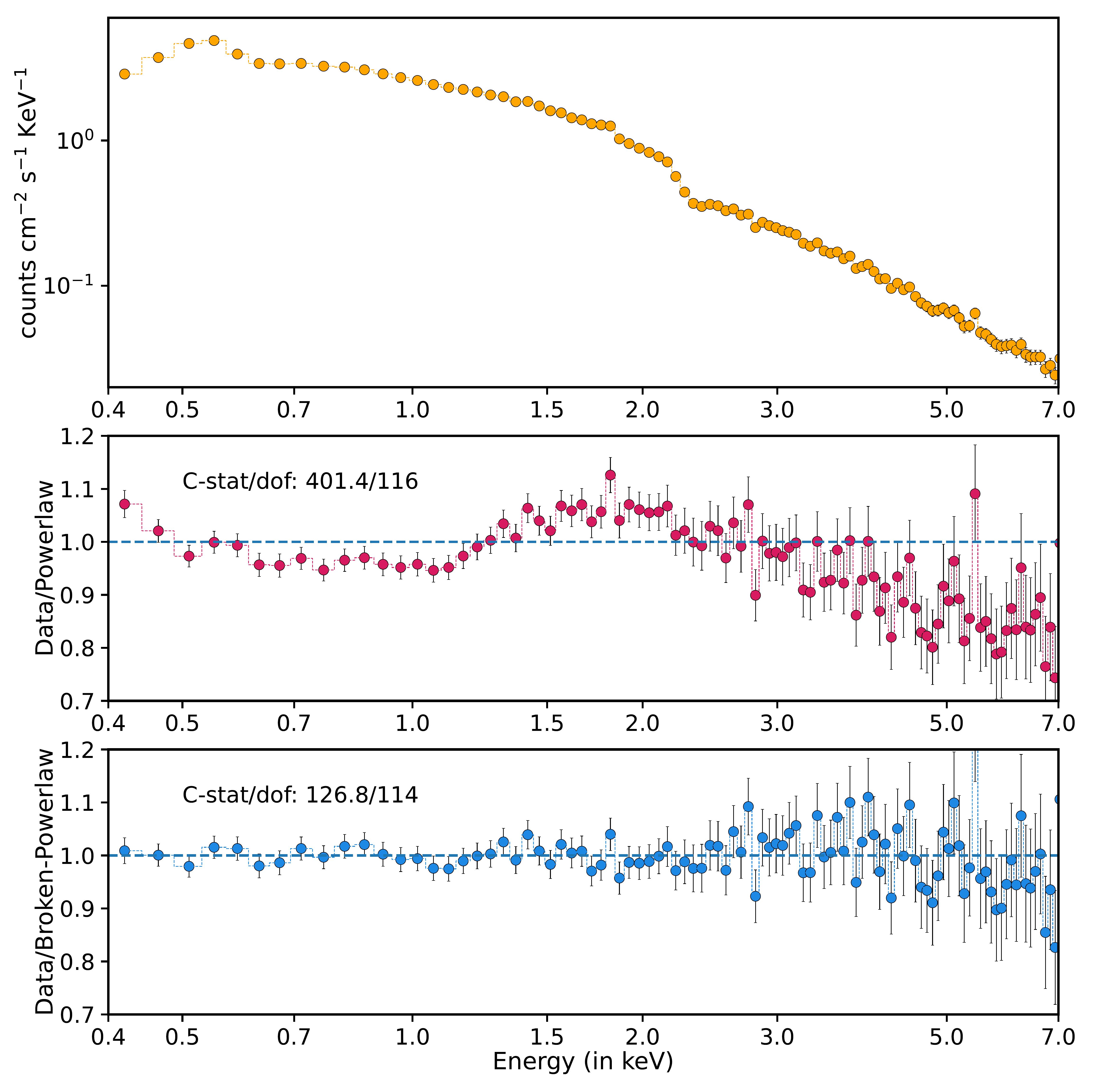}
\includegraphics[width=2.1\columnwidth]{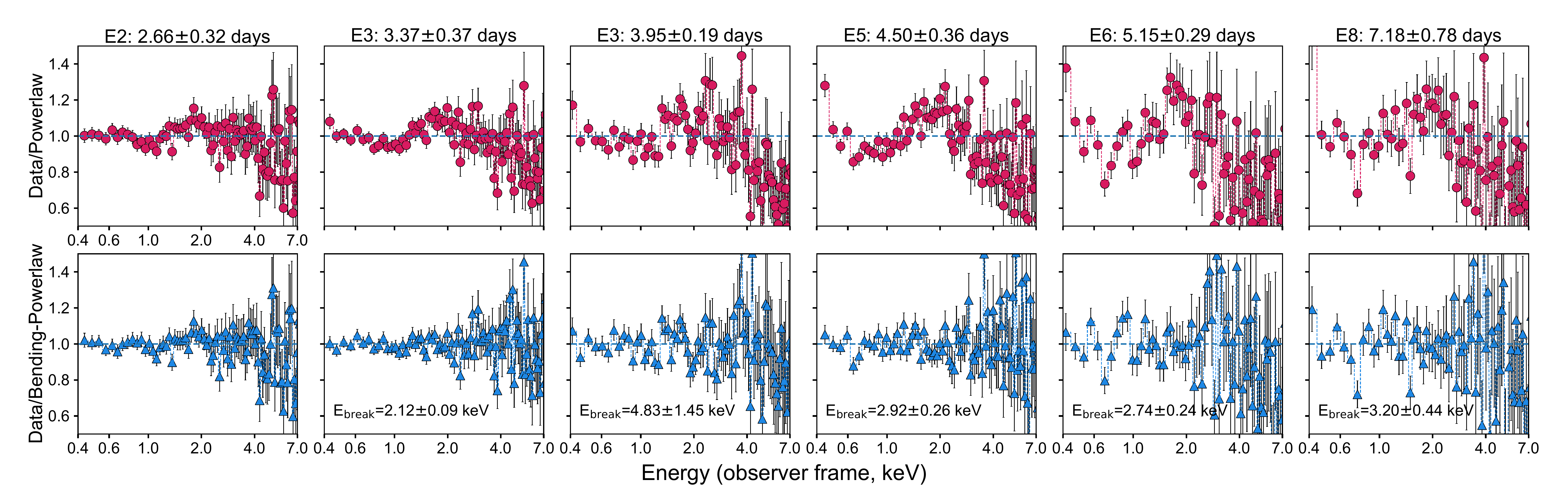}

    \caption{\textbf{Top:} Time-averaged \textit{NICER} spectrum of EP240408a using all data between MJDs 60410.567515 to 60416.699. The top panel shows the logarithm of  counts vs energy. The energy range where the source is above the \texttt{SCORPEON} background is shown, but the full $0.38$\,$-$\,$10.0$ keV was used for fitting as per \textit{NICER} data analysis guidelines. 
     \textbf{Middle:} Ratio (data to best-fit model) plots corresponding to a powerlaw and a broken powerlaw fit, respectively.  
    \textbf{Bottom:} Same as middle plots, now showing a summary of spectral modeling of early \textit{NICER} spectra. The time from EP trigger are shown at the top of each column. The top panels show the ratio of data to the best-fit powerlaw model while the bottom panels are ratios of data to best-fit broken powerlaw model. The break energy (in the rest frame) is listed in the bottom panels. These plots were generated using a redshift $z$\,$=$\,$0.5$, but the residuals are similar for $z$\,$=$\,$2.0$, though the rest frame location of the break is modified.
    }
    \label{fig:xrayspectra}
\end{figure*}

\subsubsection{\textit{Swift}/XRT}
The first X-ray observation and detection following the initial EP trigger was by \textit{Swift}/XRT approximately 1.4 d later \citep{GCN36057xrt1,ATel16585xrt2}. We retrieved the XRT spectra using the \textit{Swift}/XRT data products generator. The data were modeled in \texttt{XSPEC} \citep{Arnaud1996} using the C-statistic \citep{Cash1979}. Previous reports of the spectra \citep{GCN36057xrt1,ATel16585xrt2} assumed redshift $z$\,$=$\,$0$. For the same assumptions (a $z$\,$=$\,$0$ absorbed powerlaw; \texttt{tbabs*pow}), we confirm the prior reports of a hydrogen column density $N_\textrm{H}$\,$=$\,$(1.4\pm0.5)\times 10^{21}$ cm$^{-2}$, photon index $\Gamma$\,$=2.10\pm0.18$, and unabsorbed flux $F_\textrm{X}$\,$=$\,$(2.5^{+0.3}_{-0.2})\times10^{-11}$ erg cm$^{-2}$ s$^{-1}$ in the $0.3$\,$-$\,$10$ keV band. 

Here we model the data with an absorbed power-law model \texttt{ztbabs*zashift*(clumin*pow)} assuming either $z$\,$=$\,$0.5$ or $z$\,$=$\,$2$ due to the unknown distance (\S \ref{sec: host}). For $z$\,$=$\,$0.5$, we find $N_\textrm{H}$\,$=$\,$(3.0\pm0.6)\times 10^{21}$ cm$^{-2}$, $\Gamma$\,$=2.09\pm0.08$, and a $0.3$\,$-$\,$10$ keV rest frame X-ray luminosity of $L_X$\,$=$\,$(5.85^{+0.40}_{-0.24})\times10^{46}$ erg  s$^{-1}$. For $z$\,$=$\,$2$, we instead derive  $N_\textrm{H}$\,$=$\,$(1.1\pm0.3)\times 10^{22}$ cm$^{-2}$, $\Gamma$\,$=1.85\pm0.11$, and a $0.3$\,$-$\,$10$ keV rest frame K-corrected X-ray luminosity of $L_X$\,$=$\,$(4.8\pm0.6)\times10^{48}$ erg  s$^{-1}$. We compare this to the \textit{NICER} results below.

\subsubsection{\textit{NICER}}
\label{sec:nicerspectra}

Here we focus on the analysis of the full \textit{NICER} dataset of EP240408a. We introduced the \textit{NICER} dataset in \S \ref{sec: nicer}. Here we provide a detailed description of the data extraction and analysis, specifically with regard to the spectra. 

As per the \textit{NICER} data analysis guide, for each obsID we extracted separate "clean" event files for both the night and day time data. We used the \texttt{nicerl3-spect} tool to extract separate night and day spectra from each ObsID. We started our spectral analysis by inspecting these spectra, several of which showed excess in the $0.5$\,$-$\,$0.6$ keV band. This is a known contamination issue originating from the foreground Oxygen line complex from the Earth's atmosphere. This can happen in both the day and night time data and at present, there is no tool provided by the \textit{NICER} team to mitigate this problem. 

In order to reduce contamination from the Oxygen line we designed the following methodology. 
First, we divided the entire \textit{NICER} light curve into 18 sub-intervals whose boundaries were chosen based on the Bayesian blocks algorithm \citep{blocks} on the $0.3$\,$-$\,$10$ keV light curve extracted from \texttt{nicerl3-lc} tool. We extract both night and day time spectra from each of the 18 Bayesian blocks, resulting in 36 X-ray spectra. 

For each spectrum we applied the following procedure. We ignored spectra with less than 300 seconds of exposure and only focus on the most reliable data. If the exposure is more than 300 s, we fit the spectrum in \texttt{XSPEC} \citep{Arnaud1996}, using the C-statistic \citep{Cash1979}, with the \texttt{SCORPEON} background model. Following the recommendation of the \textit{NICER} team\footnote{\url{https://heasarc.gsfc.nasa.gov/docs/nicer/analysis_threads/cal-recommend/}}, we fit the day time and night time spectra in the $0.25$\,$-$\,$10$ keV and $0.38$\,$-$\,$10$ keV bands, respectively. We fit the spectrum twice: first allowing for the normalization of the Oxygen K foreground emission line to vary and second with the normalization fixed to 0. We then use the Akaike Information Criterion (AIC) to determine whether there is significant contamination from the Oxygen line (requiring $\Delta$AIC\,$<$\,$-20$ between the two models). 

We then fit the spectrum again using an absorbed powerlaw model \texttt{ztbabs*zashift(clumin*pow)}, including an Oxygen line if $\Delta$AIC\,$<$\,$-20$ from the previous fit.  
In order to determine whether the source is detected in an individual spectrum we require $\Delta$AIC\,$<$\,$-20$ when comparing the fit with and without the powerlaw model in addition to the \texttt{SCORPEON} background model. 
We note that automated procedures within the \texttt{NICERDAS} software (e.g., \texttt{nicerl3-lc}) detect flaring periods at later times, but we determine these are due to the Earth's atmosphere (Oxygen K emission). 
This is consistent with the deeper non-detections by \textit{Swift}/XRT over similar time periods (see Table \ref{tab: observationsXray}).

Some of the spectra yielded a reduced C-statistic of $>$2 for our powerlaw model, see the bottom panel in Figure \ref{fig:xrayspectra}. The ratio plots indicated the presence of a break in the spectrum. Thus, we followed the same methodology using $\Delta$AIC for a broken powerlaw model  (\texttt{ztbabs*zashift(clumin*bknpow)}) as well. We then compared the $\Delta$AIC between the powerlaw (\texttt{pow}) and broken powerlaw (\texttt{bknpow}) models. We find a broken powerlaw is the preferred model in epochs 3, 4, 5, 6, and 8 (referred to as, e.g., E3, E4, etc.).  Using a broken powerlaw model resulted in reduced C-statistic values close to unity (Figure \ref{fig:xrayspectra}; bottom panel), without the significant residuals observed in the powerlaw only model. This spectral break is observed in both day and night spectra.  

\begin{figure*}
    \centering
\includegraphics[width=2\columnwidth]{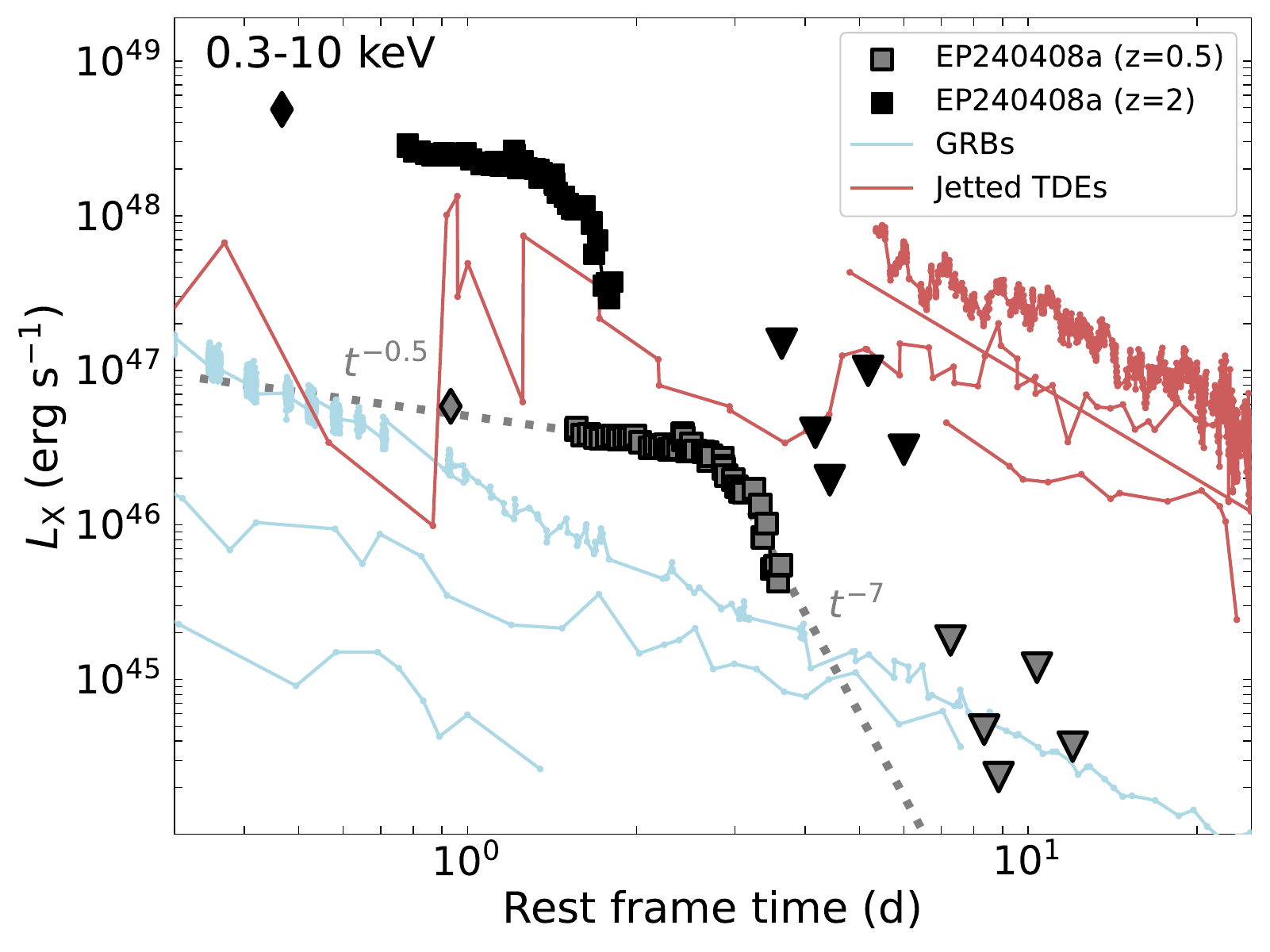}
    \caption{Comparison between the rest frame ($0.3$\,$-$\,$10$ keV; K-corrected) X-ray lightcurve of EP240408a (black) at different assumed redshifts ($z$\,$=$\,$0.5$ and $2$) versus a sample of luminous GRBs in blue (GRB 070110, GRB 221009A, and EP240315), and jetted TDEs in red \citep{Bloom2011,Levan2011,Cenko2012,Brown2015,Pasham2023}. \textit{NICER} data  of EP240408a are shown by squares, and \textit{Swift} data by a diamond for the first detection at 1.4 d and upper limits from \textit{Swift} and \textit{NuSTAR} as downward triangles. All data and limits are in the rest frame $0.3$\,$-$\,$10$ keV energy range. The gray dotted lines show a broken power-law with initial slope $t^{-0.5}$ breaking to $t^{-7}$ at $\sim$4.2 d in the observer frame. 
    Figure reproduced from \citet{Pasham2023}. 
    }
    \label{fig:xraylc}
\end{figure*}

For each spectral fit, we derived a count rate to unabsorbed K-corrected $0.3$\,$-$\,$10$ keV luminosity conversion factor. We then use this to estimate a light curve (luminosity versus time) on a per good time interval (GTI) basis. The lightcurve is shown in Figure \ref{fig:xraylc}. 

Based on our spectral analysis (for $z$\,$=$\,$0.5$), we find a typical Hydrogen column density of $N_\textrm{H}$\,$\approx$\,$(1.5-2.5)\times 10^{21}$ cm$^{-2}$, photon index $\Gamma_1$\,$=1.5$\,$-$\,$2.0$, and an observer frame spectral break at $E_\textrm{break}$\,$\approx$\,$4.5$ keV where the spectra steepens to $\Gamma_2$\,$=2.5$\,$-$\,$3.0$. The evolution of the spectral index is rather unconstrained by the data, due to the decreasing source count statistics at higher energies and a variable background rate. The photon index $\Gamma_1$ and column density $N_\textrm{H}$ are comparable to that derived from the initial \textit{Swift}/XRT detection. There is marginal evidence that the photon index $\Gamma_1$ decreases from $\sim$\,$2$ to $\sim$\,$1.5$ over the first few days. 

The \textit{NICER} spectra also revealed the requirement of an intrinsic absorption component in excess of the Galactic value of $N_\textrm{H,gal}$\,$=$\,$8.5\times 10^{20}$ cm$^{-2}$ along the line-of-sight \citep{Willingale2013}. The expected optical extinction $A_V$ from the inferred hydrogen column density $N_\textrm{H}$\,$\approx$\,$(1.5-2.5)\times 10^{21}$ cm$^{-2}$ is $A_V$\,$\approx$\,$1$ mag at $z$\,$\approx$\,$0.5$ \citep{Guver2009}, with larger values required at higher redshifts (Appendix \ref{sec:xrayspectra}).

\subsection{\textit{NICER} X-ray Lightcurve}
\label{sec:xraylightcurve}

\textit{NICER} observed \target at high-cadence between 1.8 and 38 d post trigger. The overall light curve shows a slow decline over the first few days followed by a rapid drop off (see Figure \ref{fig:xraylc}). The count rate lightcurve decays from $\sim$\,$8$ cts s$^{-1}$ to $\sim$\,$5$ cts s$^{-1}$ over $\sim$\,$2$ d before steepening and rapidly fading below detection over the course of $\sim$\,$1$ day. We can model the count rate lightcurve with a broken powerlaw of $t^{-0.5}$ breaking to $t^{-5}$ around $\sim$4.1 to 4.3 d after the EP trigger. This steep break is seen in both day and night time data. 

However, the luminosity lightcurve (see \S \ref{sec:data}) is significantly steeper ($t^{-7}$; Figure \ref{fig:xraylc}) as it properly accounts for the background, in particular the Oxygen line (\S \ref{sec:nicerspectra}). 
We used the \texttt{dynesty} \citep{dynesty} nested sampling package to fit the X-ray ($0.3$\,$-$\,$10.0$ keV) flux lightcurve with a broken powerlaw which yields a slope of $t^{-0.51\pm0.05}$ early-on, which breaks to $t^{-6.7\pm0.2}$ around an observer frame time of $4.19\pm0.02$ d post-trigger. We infer a timescale $\delta \tau/\tau$\,$\approx$\,$0.2$ for the plateau's decline. 

We briefly consider other possible explosion times for EP240408a, prior to the EP trigger. If we allow for an explosion (or disruption; depending on interpretation) time of $\sim$\,$3$ d prior to the trigger the lightcurve can be roughly modeled with a broken power-law of $t^{-5/3}$ and $t^{-15}$. If taken to be $10$ d prior, these slopes are instead $t^{-2.2}$ and $t^{-25}$ with the break occurring at $\sim$\,$14.1$\,$-$\,$14.3$ d (observer frame). It is possible for EP to miss the X-ray transient in prior observations accounting for natural variability of the X-ray lightcurve as shown by Sw J1644+57 at early times. However, the significant difficulty in this case is explaining the EP trigger as a $\sim$\,$1,000\times$ brighter flare at $\sim$\,$3$\,$-$\,$10$ d after the initial onset.

\section{Discussion} \label{sec: discuss}

\subsection{An Optical Search Reveals A Potential Host}\label{sec: host}

Our deep Gemini imaging (\S \ref{sec: gemini}) uncovers a candidate host galaxy at the South-East edge of the updated XRT enhanced localization (Figure \ref{fig:fc}). We detect this source only in our optical $r$-band and $i$-band images with brightness $r$\,$\approx$\,$24$ AB mag (Table \ref{tab: observationsOpt}). The source is not detected in our near-infrared imaging to depth $J$\,$\gtrsim$\,$23.2$ AB mag. As we detect this source only in two filters, no robust constraints on its spectral energy distribution or distance scale can be obtained, and there is no existing optical or near-infrared spectroscopy of this source. We performed difference imaging between thee Gemini observation epochs with the Saccadic Fast Fourier Transform (\texttt{SFFT}) software \citep{Hu2022}, but did not identify any optical or near-infrared variability near either the initial position, enhanced XRT position, or the location of this candidate host (\S \ref{sec: gemini} and Table \ref{tab: observationsOpt}).

The probability of chance coincidence $P_\textrm{cc}$ of the candidate host with the $3\sigma$ XRT localization is $P_\textrm{cc}$\,$\approx$\,$0.15$ \citep{Bloom2002,Berger2010}. This is generally considered an inconclusive association for an extragalactic transient. A typical cutoff for a conclusive association is $P_\textrm{cc}$\,$<$\,$0.1$ \citep[e.g.,][]{Berger2010,Fong2013,Fong2022,OConnor2022}. In this case, the lack of a subarcsecond localization prevents a robust determination, but we identify this source as the most likely host for EP240408a among the sources we are able to resolve in our deep optical and near-infrared images. If we expand our search for other potential host galaxies to all sources within $\sim90\arcsec$ following the methods outlined in \citet{OConnor2022}, we find 6 clearly extended galaxies within this region with the closest offsets of $25$\,$-$\,$26$\arcsec. We compute the probability of chance coincidence $P_\textrm{cc}$ \citep{Bloom2002,Berger2010a} for these galaxies using their $r$-band apparent after correcting for Galactic extinction $E(B-V)$\,$=$\,$0.076$ mag \citep{Schlafly2011}. Each of these galaxies has $P_\textrm{cc}$\,$>$\,$0.6$\,$-$\,$0.95$ and all are unlikely hosts to \target.
We therefore conclude that EP240408a is extragalactic in nature and likely related to the faint galaxy lying at the edge of the XRT localization (Figure \ref{fig:fc}).

From Gemini we have deep upper limits on any other underlying source within the XRT localization of $r$\,$>$\,$26$, $i$\,$>$\,$26.1$, and $J$\,$>$\,$23.2$ AB mag. These limits are capable of excluding the majority of GRB \citep[e.g.,][]{Palmerio2019,Schneider2022,OConnor2022,Nugent2022}, and TDE \citep{Hammerstein2021} host galaxies out to $z$\,$\sim$\,$0.5$\,$-$\,$1$, and imply a higher redshift for EP240408a. The brightness of our candidate host galaxy, when compared to other high-energy extragalactic transient host galaxies, also potentially places it at higher redshifts $z$\,$\gtrsim$\,$1$ in the range of $0.1$\,$-$\,$1.0L_*$ galaxies \citep[see, e.g.,][]{OConnor2022}. 
However, we cannot exclude that it is a small galaxy ($10^{8-9} M_\odot$) at $z$\,$\approx$\,$0.5$. 
For example, adopting galaxy correlations between the $i$-band absolute magnitude and stellar mass \citep{Taylor2020}, our observations are sensitive to a $10^8 M_\odot$ galaxy out to $z$\,$\approx$\,$0.5$ and a $10^9 M_\odot$ galaxy out to $z$\,$\approx$\,$2$. Future optical spectroscopy of this source can constrain its distance scale, though given its brightness such observations are challenging. In what follows (\S \ref{sec:extscen} and \ref{sec:interp}) we treat the distance scale of EP240408a as unconstrained and consider redshifts between $z$\,$\approx$\,$0.5$\,$-$\,$2$. Lower redshifts are strongly disfavored by the lack of bright optical, near-infrared, or radio counterparts.

\subsection{Disfavoring a Galactic Origin}
\label{sec:galactic} 

Here we argue that the observed brightness of the initial EP trigger strongly disfavors a Galactic nature for the source as the luminosity and other properties are inconsistent with Galactic transients such as cataclysmic variables (CVs), high-mass X-ray binaries (HMXBs), low-mass X-ray binaries, and Galactic magnetars.

The lack of a bright optical counterpart, evidence for a hydrogen column density larger than the line-of-sight Galactic value (see \S \ref{sec:nicerspectra}), lack of X-ray periodicity \citep[Appendix  \ref{sec:timing} and][]{ATel16589saltnicer}, and smooth decline of the lightcurve lacking any short timescale X-ray variability, all strongly disfavor accreting binaries such as CVs, HMXBs, or LMXBs. The X-ray luminosity of the source at 20 kpc approaches $L_X$\,$\approx$\,$10^{39}$ erg s$^{-1}$ ($0.5$\,$-$\,$4$ keV) which would be at the high end of the luminosity function for XRBs, though not completely unreasonable. One reason to consider a Galactic scenario is that EP240408a is located at a relatively low Galactic latitude with Galactic coordinates $l$, $b$ = 274.16 deg, 19.41 deg. However, as there is almost no Galactic dust in this field ($A_V$\,$\approx$\,$0.2$ mag; \citealt{Schlafly2011}), which could hide luminous Galactic objects, we should have detected a high mass star in our Gemini images out to even $>$\,$100$ kpc, and M dwarf stars should be detected out to 15 kpc even in quiescence.

The lack of X-ray periodicity (Appendix  \ref{sec:timing}) and lack of a low temperature X-ray spectrum (or blackbody excess) can also be used to strongly disfavor a Galactic magnetar as the origin of EP240408a. The archival X-ray limit on a quiescent X-ray flux from \textit{eROSITA} \citep{erositaULs} assuming a distance of $1$\,$-$\,$10$ kpc is $\lesssim$\,$10^{31-33}$ erg s$^{-1}$ ($0.2$\,$-$\,$6$ keV). For comparison the observed quiescent X-ray luminosity for Galactic magnetars is in the range $10^{30-35}$ erg s$^{-1}$ \citep{CotiZelati2017}. 

An extensive literature search for similar lightcurves revealed that the X-ray outbursts of accreting millisecond X-ray pulsars (MSPs) display comparable behavior \citep{Wijnands2004}. Their outbursts can be modeled by multiple exponential decays defined by $e^{-t/\tau}$, where the initial exponential decay timescale $\tau$ is typically on the order of $5$\,$-$\,$10$ d and can steepen to a decay timescale of $1$\,$-$\,$2$ d \citep{Gilfanov1998,Markwardt2002,Falanga2005,Jonker2010,Sanna2018}. The break time is usually around $10$\,$-$\,$20$ d after discovery. 
We find \target's X-ray luminosity lightcurve (Figure \ref{fig:xraylc}) can be fit with $e^{-t/6}$ breaking to $e^{-t/0.8}$ after $\sim$\,$4$ d. 
However, the major difference between transient MSP outbursts and EP240408a lies in the X-ray spectra. The MSP outbursts display spectral cutoff energies in excess of 100 keV, significantly higher than the $\sim$\,$4$ keV spectral break observed by \textit{NICER} (Figure \ref{fig:xrayspectra}). We also found no evidence for periodicity in the X-ray data, even performing an accelerated pulsar search (see Appendix  \ref{sec:timing}). For these reasons, we disfavor a MSP as the explanation for EP240408a. If EP240408a is Galactic in origin, which we disfavor, it would represent a peculiar, and potentially unique, X-ray transient.

\subsection{Extragalactic Scenarios for EP240408a}
\label{sec:extscen}

We consider a number of possible extragalactic scenarios capable of explaining \target's observed properties (high X-ray luminosity, sharp X-ray decay,  and lack of optical or radio counterpart to deep limits). 

\begin{figure*}
    \centering
\includegraphics[width=1\columnwidth]{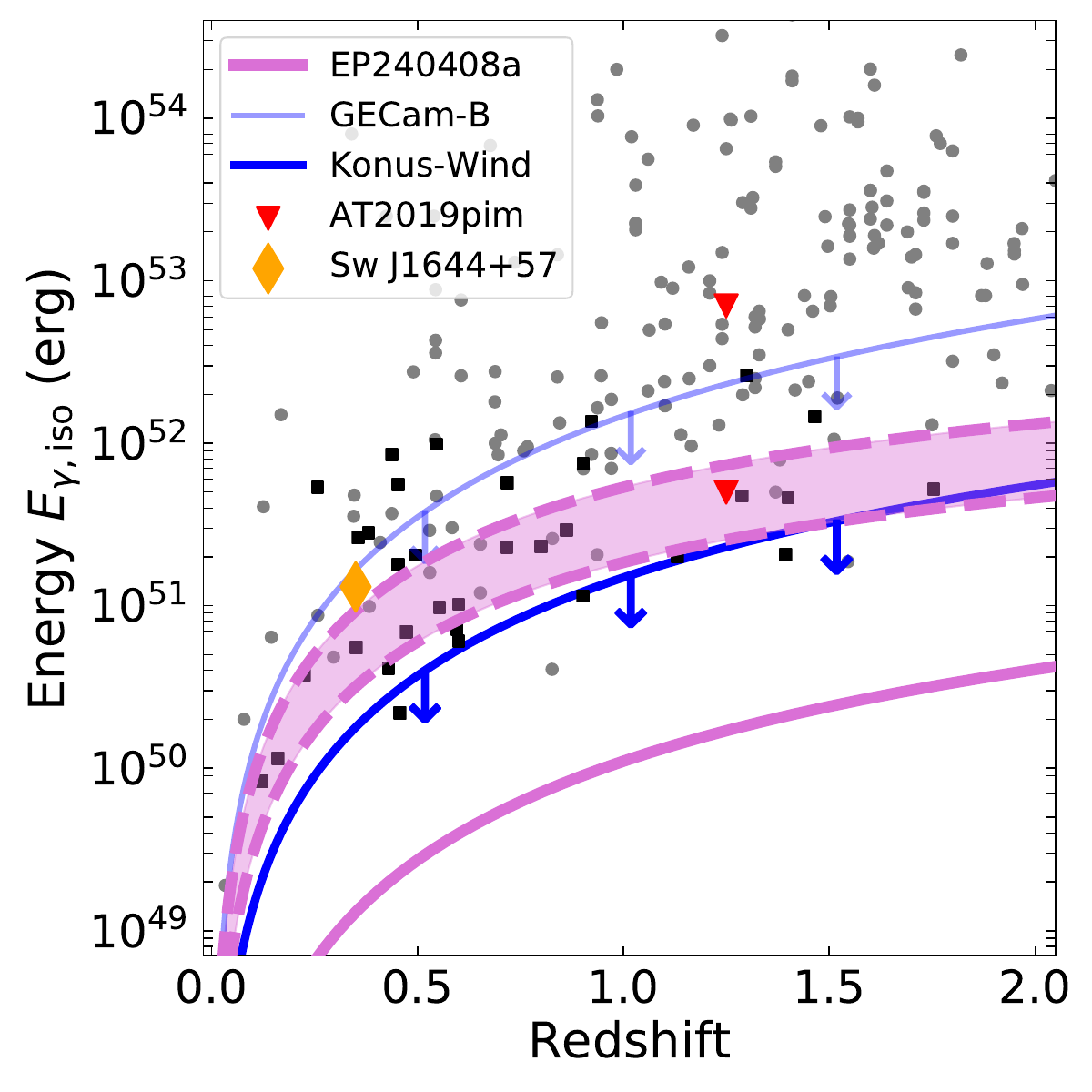}
\includegraphics[width=1\columnwidth]{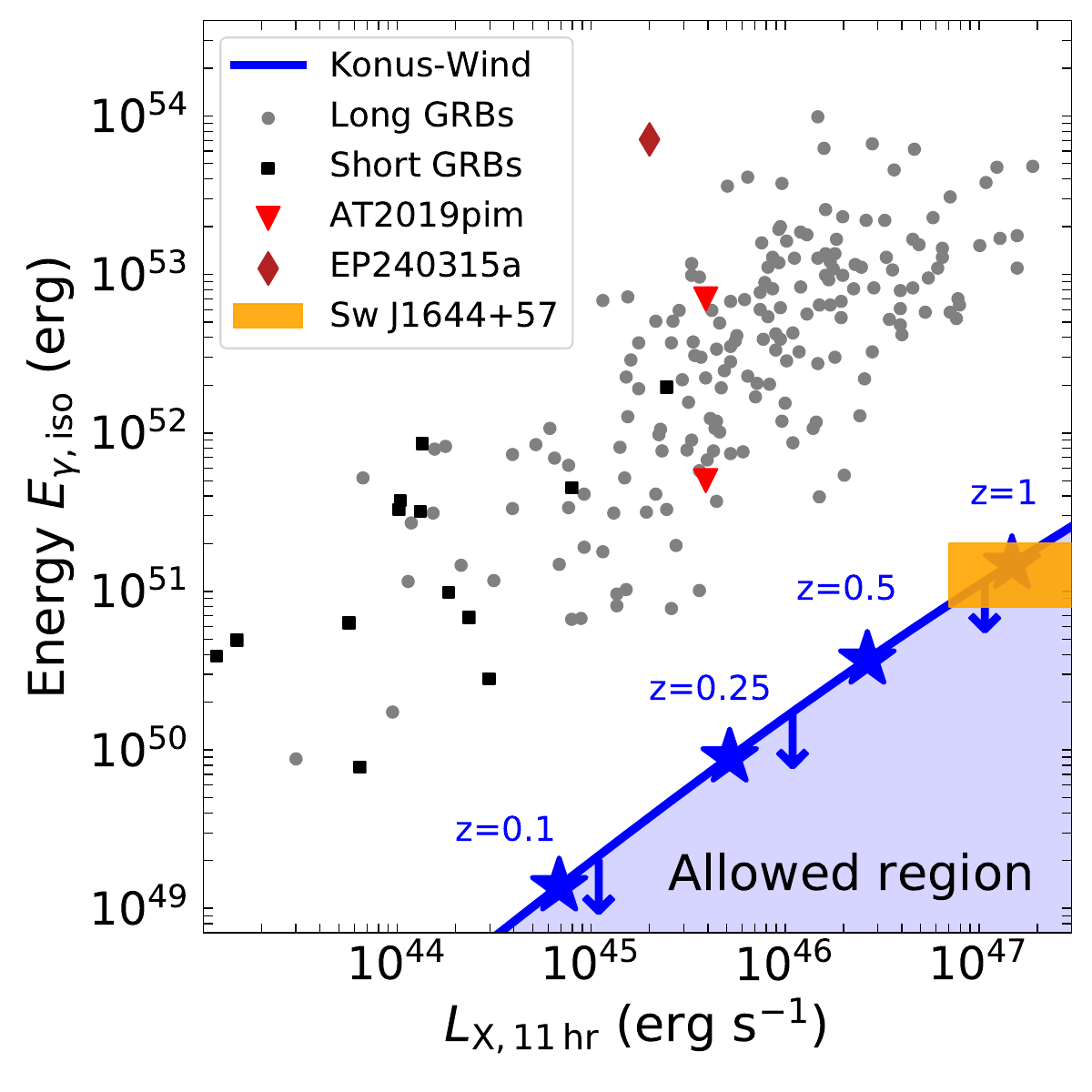}
    \caption{\textbf{Left:} Distribution of isotropic-equivalent gamma-ray energies ($10$\,$-$\,$10,000$ keV) of both short (black squares) and long duration (gray circles) GRBs versus redshift \citep{Sakamoto2011,Lien2016,Atteia2017,OConnor2024}. The approximate energy ($0.5$\,$-$\,$4$ keV) of the initial EP detection of EP240408a is shown by a thick purple line. 
    Adopting a typical range of long GRB prompt emission spectra ($E_\textrm{p}$\,$=$\,$70$\,$-$\,$300$ keV) we estimate the $10$\,$-$\,$10,000$ keV gamma-ray energy of EP240408a (shaded region between purple dashed lines). This is compared to upper limits on the gamma-ray energy from \textit{GECam-B} (solid light blue line) and \textit{Konus-Wind} (solid dark blue line; Dmitry Svinkin, private communication). 
    Reproduced from \citet{OConnor2023}.
    \textbf{Right:} 
    Rest frame X-ray luminosity ($0.3$\,$-$\,$10$ keV) at 11 hours versus gamma-ray energy ($10$\,$-$\,$10,000$ keV). A sample of short (black squares) and long GRBs (gray circles) are compiled from \citet{Nysewander2009,Berger2014}. Other strange gamma-ray bursts such as EP240315a \citep{Levan2024,Liu2024} and AT2019pim \citep{Perley2024} are also shown. We also display the approximate location of Sw J1644+57 \citep{Bloom2011,Levan2011,Burrows2011} using the energy of the initial prompt gamma-ray trigger \citep{Sakamoto2011J1644} converted to $10$\,$-$\,$10,000$ keV (see text). 
    The blue stars mark multiple redshifts up to $z$\,$=$\,$1$, higher redshifts produce further tension with observed long GRBs. The solid blue line represents the luminosity of the \textit{Swift}/XRT detection of EP240408a at 1.4 d (observer frame). 
    }
    \label{fig:grb}
\end{figure*}

\subsubsection{Constraints on Prompt Gamma-ray Emission}
\label{sec:prompt}

As EP240408a is likely a luminous extragalactic transient (peak X-ray luminosity $L_X$\,$\approx$\,$10^{49}$ erg s$^{-1}$ at $z$\,$=$\,$0.5$; $0.5$\,$-$\,$4$ keV), the natural first interpretation is a long duration GRB. This is simply due to the frequent observed rate of long GRBs, approximately one per day detected by all-sky monitors such as \textit{Fermi} and \textit{Swift}, and the similar prompt timescale of $\sim$10 s. Moreover, since the launch of EP there have been a number of likely GRBs that lack hard gamma-ray detections \citep[e.g.,][]{Yin2024}, though we note that given the soft X-ray trigger these are not effectively true ``orphan'' GRBs \citep{Nakar2002,Huang2002,Dalal2002}. 

In Figure \ref{fig:grb} (left panel), we compare the constraints on the prompt gamma-ray energy (isotropic-equivalent) from \textit{GECam-B} and \textit{Konus-Wind} to a population of long and short GRBs. We also show the estimated gamma-ray energy ($10$\,$-$\,$10,000$ keV) for EP240408a adopting a typical range of long GRB prompt emission spectra ($E_\textrm{p}$\,$=$\,$70$\,$-$\,$300$ keV; $\alpha$\,$=$\,$-1$; $\beta$\,$=$\,$-3$; \citealt{Band1993}). 
This comparison shows that there exist a variety of short and long GRBs consistent with the \textit{GECam-B} limits, but that the expected \textit{Konus-Wind} limits are only marginally consistent with the expected prompt emission of EP240408a (if interpreted as a GRB). 

A more informative diagnostic comes from combining limits on the prompt emission with observations of the X-ray afterglow. A standard diagnostic is the ratio of the X-ray flux at 11 hours $F_\textrm{X,11}$ to the gamma-ray fluence $\phi_\gamma$ \citep{Nysewander2009,Berger2014,OConnor2020,OConnor2022,Yang2024,Perley2024}, which is approximately distance independent with only a slight correction for redshift \citep[see][for a discussion]{OConnor2020}. We find that E240408a has a high ratio $F_\textrm{X,11}/\phi_\gamma$\,$>$\,$10^{-5}$, already higher than the most luminous GRBs \citep[e.g.,][]{Nysewander2009,Berger2014}. This implies that the X-rays are very bright compared to any possible gamma-ray emission. We further show this in Figure \ref{fig:grb} (right) which displays the X-ray luminosity at rest frame 11 hours $L_\textrm{X,11}$ versus the isotropic-equivalent gamma-ray energy. The allowed space for EP204048a (blue shaded region) is clearly separate  from any short or long GRB at all redshifts, though the difference increases at $z$\,$>$\,$1$.  
We note that this comparison would become even more significant if the X-ray luminosity at $>$\,$1$ day was compared instead of the earlier time of 11 hours as the majority of GRBs decay as $t^{-1}$ whereas EP240408a displays a slower fading plateau-like decay.

The constraints on EP240408a (Figure \ref{fig:grb}) push it towards the region occupied by the relativistic jetted TDE Sw J1644+57 \citep{Bloom2011,Levan2011}. We converted the $15$\,$-$\,$150$ keV fluence of the initial BAT trigger \citep{Sakamoto2011J1644} of Sw J1644+57 to the $10$\,$-$\,$10,000$ keV energy range assuming a peak energy of $E_\textrm{p}$\,$\approx$\,$70$ keV \citep{Levan2011}. 
We further consider the jetted TDE scenario in \S \ref{sec:tde}.

\begin{figure*}
    \centering
\includegraphics[width=\columnwidth]{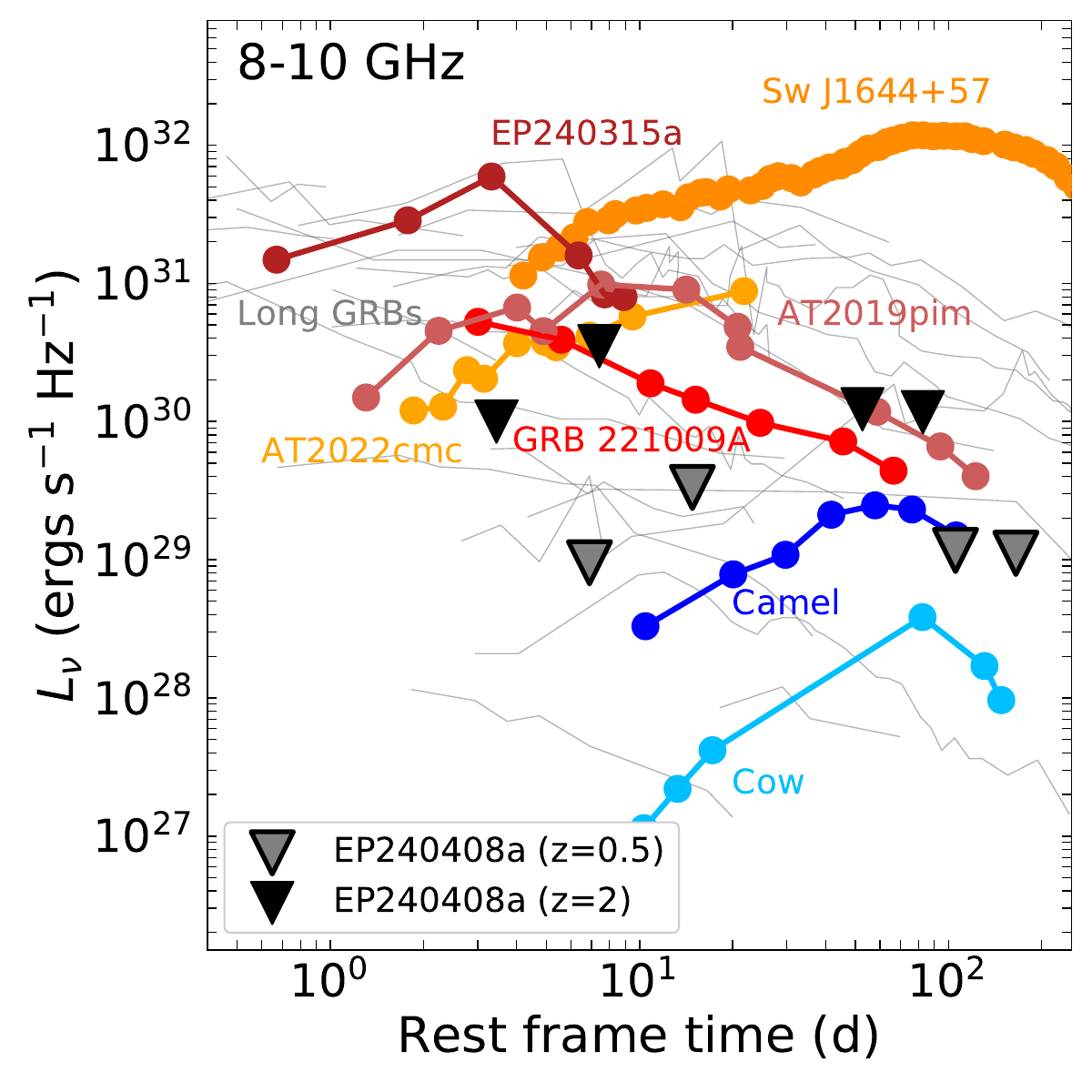}
\hspace{5mm}
\includegraphics[width=\columnwidth]{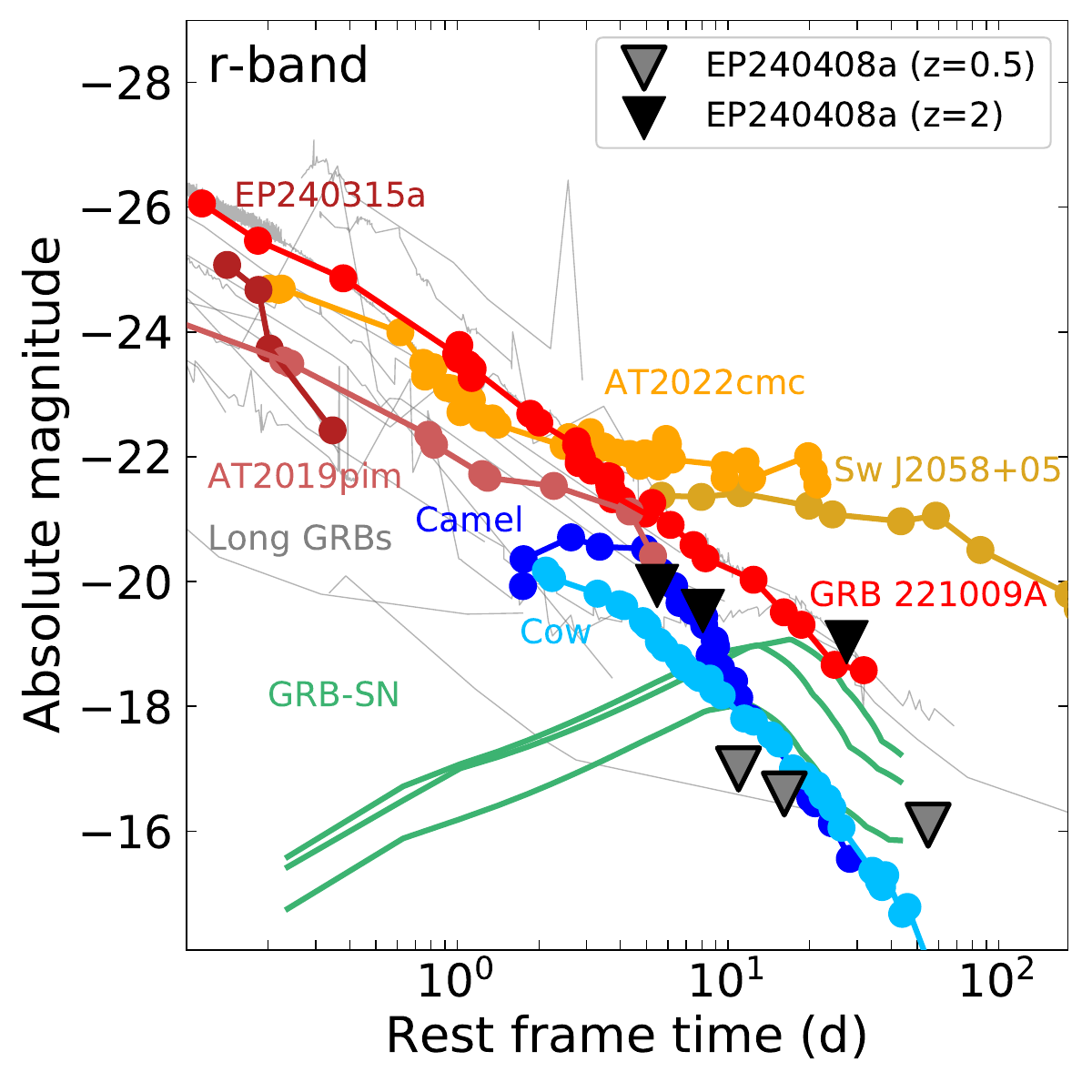}
    \caption{\textbf{Left:} Comparison between the 10 GHz upper limits for EP240408a (downward triangles) at different assumed redshifts versus the rest frame radio luminosity of multiple classes of energetic transients, including long duration GRBs \citep{ChandraFrail2012,Laskar2022}, jetted TDEs \citep{Zauderer2011,Andreoni2022}, and FBOTs \citep{Ho2019cow,Margutti2019cow,HoKoala,Coppejans2020cow}. EP240315a was compiled from \citet{Ricci2024} and AT2019pim from \citet{Perley2024}. 
    \textbf{Right:} A similar comparison for observed $r$-band absolute magnitudes of various energetic transients: FBOTs \citep{Perley2019,Margutti2019cow,PerleyCamel}, TDEs \citep{Cenko2012,Pasham2015,Andreoni2022,Pasham2023}, GRB-SN \citep{Galama1998,Srinivasaragavan2023,Srinivasaragavan2024}, GRB 221009A \citep{OConnor2023}, AT2019pim \citep{Perley2024}, and other GRB afterglows \citep{Dainotti2024}. The thin gray lines are long GRBs. The upper limits from our deep Gemini GMOS-S observations are shown as downward triangles at $z$\,$=$\,$0.5$ and $z$\,$=$\,$2$. }
    \label{fig:ULs}
\end{figure*}

\subsubsection{The Multi-wavelength Properties of EP240408a}

Here we briefly compare the observed multi-wavelength properties of EP240408a to other classes of high-energy extragalactic transients. A more detailed discussion of each transient class is presented in \S \ref{sec:interp}. We focus here on a few possible candidates including GRBs, jetted TDEs, and FBOTs. 

Figure \ref{fig:xraylc} compares the rest frame X-ray ($0.3$\,$-$\,$10$ keV; K-corrected) lightcurve of EP240408a (assuming either redshifts $z$\,$\approx$\,$0.5$ or $2.0$) to GRBs and jetted TDEs. 
While the peak X-ray luminosity of GRBs can be orders of magnitude higher ($>$\,$10^{50}$ erg s$^{-1}$), jetted TDEs clearly separate themselves from long GRBs in their luminous, long lasting X-ray emission. The X-ray emission of EP240408a is also more long-lived than a typical GRB, and for a redshift $z$\,$\approx$\,$1$\,$-$\,$3$ appears to be consistent with the X-ray luminosity of jetted TDEs. We also note that the observed X-ray spectral break (Figure \ref{fig:xrayspectra}) is characteristic of non-thermal emission, and observed in both GRBs and jetted TDEs \citep[e.g., AT2022cmc;][]{Yao2024}, see \S \ref{sec:specbreak} for further discussion of this point. 

While we did not detect EP240408a in deep optical, near-infrared, or radio observations we can still compare these limits (focusing again on $z$\,$\approx$\,$0.5$ or $2.0$) to these classes of transients. In Figure \ref{fig:ULs} (right panel) we compare our Gemini limits to a variety of optical transients, finding that we can exclude most classes (e.g., GRBs, FBOTS, TDEs) out to $z$\,$\sim$\,$0.5$ (and beyond), and significantly disfavoring a low redshift $z$\,$<$\,$0.5$ for any of these classes. In addition, for $z$\,$<$\,$0.5$ the inferred hydrogen column density predicts a decreasing intrinsic extinction $A_V$\,$<$\,$1.4$ mag with decreasing redshift, making luminous optical emission even less likely to be missed. This further motivates considering higher redshift interpretations for EP240408a (see also \S \ref{sec: host}). 

In Figure \ref{fig:ULs} (left panel) we show the rest frame radio lightcurves of multiple well-known transients to our VLA and ATCA upper limits. Both jetted TDEs and long GRBs produce long-lasting luminous radio emission that should likely be detected out to $z$\,$\approx$\,$2$. We discuss potential reasons for delayed radio emission from a jetted TDE in \S \ref{sec:tde}. We also explore the implications of a non-detection on the jet's kinetic energy and the density of the surrounding environment in \S \ref{sec:grb}.


\subsection{Disfavored Extragalactic Interpretations of EP240408a}
\label{sec:interp}

Here we consider a few classes of extragalactic high-energy transients that we disfavor as the progenitor of EP240408a. In \S \ref{sec:grb} and \S \ref{sec:tde} we focus on two more likely interpretations in GRBs and TDEs, respectively. 

\subsubsection{Active Galactic Nuclei (AGN)}
\label{sec:agn}

The most luminous X-ray bright AGN have luminosities as high as $L_\textrm{X}$\,$=$\,$10^{46}$ erg s$^{-1}$ \citep{Donnarumma2009,Pineau2011}, and generally display X-ray variability on timescales of hours to days \citep{Uttley1999,Uttley2005}. The short timescale (10 s) and high luminosity for EP240408a at $z$\,$>$\,$0.5$ clearly disfavor the interpretation of the source as an AGN. A more distant source ($z$\,$>$\,$0.5$) would lead to X-ray luminosities higher than the most luminous X-ray bright AGN. 
Moreover, the most X-ray luminous AGN also display bright optical luminosities of $\approx$\,$-28$ to $-30$ AB mag which would be detected by Gemini at any reasonable distance. We therefore strongly disfavor an AGN flare as the cause of EP240408a.

\subsubsection{Fast Blue Optical Transient (FBOT)}

FBOTs are characterized by a rapidly fading ($\sim$\,$0.3$ mag d$^{-1}$), blue ($g$\,$-$\,$r$\,$\lesssim$\,$0.2$ mag) optical transient \citep[e.g.,][]{Drout2014,Pursiainen2018} with comparable absolute magnitudes ($\approx$\,$-20$ mag) to superluminous supernovae (SLSN). Following the discovery of AT2018cow \citep{Prentice2018,Margutti2019cow,Perley2019}, the canonical FBOT, they are now commonly identified in wide-field optical surveys. They have also been found with luminous X-ray emission \citep{Prentice2018,Margutti2019cow,Perley2019,PerleyCamel,Ho2019cow,Ho2020,Coppejans2020cow,Yao2022,Ho2023}. In the case of AT2018cow, the peak X-ray luminosity was $L_\textrm{X}$\,$=$\,$10^{43}$ erg s$^{-1}$ and displayed slow fading emission $t^{-1}$ until $\sim$20 d (rest frame) before a sharp decay $t^{-4}$ \citep{Margutti2019cow}. While the characteristic timescale is longer than the plateau-like emission observed from EP240408a, the behavior of a sharp break after a standard temporal decline is similar. Due to selection effects related to detecting a similar prompt X-ray transient to that identified for EP204048a, it cannot be excluded that FBOTs display similar features. Therefore we compare the luminosity of the X-ray plateau phase of EP240408a to the peak flux of AT2018cow. EP240408a would exceed AT2018cow\footnote{Another X-ray luminous FBOT, AT2024qfm, was recently discovered with luminosity as high as $8\times10^{43}$ erg s$^{-1}$ \citep{Margutti2024TNSAN_AT2024qfm}.} for $z$\,$>$\,$0.015$ ($\sim$\,$60$ Mpc), similar to the actual distance to AT2018cow, where a bright host galaxy would be visible and a large amount of dust would be required to miss a luminous cow-like optical transient in our deep, multi-epoch $riJ$ images (Figure \ref{fig:fc} and Table \ref{tab: observationsOpt}). This does not fit with the inferred intrinsic $A_V$ from the X-ray spectrum, which at such low redshifts is almost negligible compared to the expected brightness of a typical FBOT. In fact, the strongest constraint on an FBOT-like explosion lies in the lack of bright optical emission, which implies a higher redshift ($z$\,$\gtrsim$\,$0.5$) where the X-rays are orders of magnitude larger than from observed FBOTs. 
We therefore strongly disfavor an FBOT-like transient to explain EP240408a.

\subsubsection{Fast X-ray Transient (FXT)}
\label{sec:fxt}

The last decade has seen the detection of a new class of X-ray transients discovered serendipitously in deep X-ray imaging, largely with \textit{Chandra} and \textit{XMM-Newton} \citep{Jonker2013,Glennie2015,Bauer2017,Xue2019,Alp2020,Ai2021,Sarin2021,Quirola2022,Quirola2023,Quirola2024}.
Referred to as Fast X-ray Transients (FXTs) they are characterized by short-lived ($100$\,$-$\,$10,000$ s) X-ray emission with typical peak X-ray luminosities in the range $10^{44-47}$ erg s$^{-1}$ \citep{Wichern2024}. The low luminosity and short timescales of some of these events have led them to be characterized as supernova shock breakout emission \citep{Alp2020} or stellar flares \citep{Glennie2015}, while some display similarities (such as plateaus) to short GRBs, potentially linking them to binary neutron star mergers \citep[e.g,][]{Ai2021,Sarin2021,Eappachen2023,Quirola2024}. 

We note that not only are the timescales of the X-ray detections of EP240408a significantly longer than known FXTs (Figure \ref{fig:grbplat}), EP240408a has a peak X-ray flux two to three orders of magnitude higher than the brightest FXTs, which if drawn from the same population should imply a low redshift with an obvious host galaxy association. For example, CDF-S XT1 \citep{Bauer2017} is associated to an extremely faint host galaxy in deep \textit{HST} and \textit{JWST} imaging with a photometric redshift of $z$\,$\approx$\,$2.8$ \citep{Quirola2024host}. Even at this large distance, the peak X-ray luminosity is only $3\times10^{47}$ erg s$^{-1}$ \citep{Quirola2024host}. 
In contrast, already at $z$\,$\approx$\,$0.1$ ($\sim$\,$460$ Mpc)  the peak X-ray luminosity ($>$\,$10^{47}$ erg s$^{-1}$) of EP240408a already exceeds that of the most luminous FXTs making this interpretation unlikely.

\subsection{A Peculiar Gamma-ray Burst}
\label{sec:grb}

GRBs are extreme transients with isotropic-equivalent energies in the range of $10^{48-55}$ erg. They are thought to be produced by either the merger of compact objects \citep{Abbott+17-GW170817A-MMO,Savchenko2017,Goldstein2017} or the collapse of massive stars \citep[collapsars;][]{Woosley1993,Macfadyen1999}. 
Historically, GRBs have been separated into the classes of short versus long duration GRBs based on the duration of their prompt gamma-ray emission with the divide at either shorter or longer than 2 seconds \citep{Kouveliotou1993}. However, there is growing evidence \citep[e.g., GRBs 060614, 211211A and 230307A;][]{DellaValle2006,Galyam2006,Yang2015,Rastinejad2022,Troja2022,Yang2022kn211211A,Gompertz2023,Levan2023,Yang2024,Gillanders2023,Dichiara2023} that the duration separation is not a robust classifier of the GRB's progenitor. Therefore, despite the $\sim$\,$10$ s duration of the ``prompt'' phase of EP240408a we do not automatically exclude the merger of two compact objects (e.g., two neutron stars) as its progenitor. 

\subsubsection{A ``Naked'' GRB}

The prompt phase of GRBs is generally shortly followed by a multi-wavelength ``afterglow'' from the forward shock (FS) produced by the interaction of the relativistic jet and the surrounding environment \citep{Meszaros1997,Sari1998,Wijers1999}. Standard GRB afterglows at optical and X-ray wavelengths are characterized by a decay of $t^{-1}$ followed by a $t^{-2}$ break due to the collimation of the outflow \citep{Rhoads1999,SariPiranHalpern1999,Frail2001}. This afterglow phase is not observed in the X-ray lightcurve of EP240408a, and instead we identify a plateau followed by a steep decline. 

A significant fraction of both short and long GRBs also display X-ray plateaus in their early lightcurves, many of which are followed by extremely fast decays (even as steep as $t^{-9}$; \citealt{Troja2007}) that are similar to EP240408a, which shows $\delta \tau/\tau$\,$\approx$\,$0.2$. The standard FS emission is incapable of decays steeper than $t^{-p}$ \citep[e.g.,][]{SariPiranHalpern1999}, where $p$ is the slope of the electron's powerlaw energy distribution, or, in the ``best case scenario'' $t^{-(2+\beta)}$\,$\approx$\,$t^{-3}$ if the source somehow stops emitting abruptly and only high-latitude emission is observed \citep{KumarPanaitescu2000}. Therefore, plateaus followed by such steep declines, referred to as ``internal plateaus'' \citep[e.g.,][]{Zhang2006,Troja2007}, are generally interpreted as being due to long-lived central engine activity. 

Provided the observed X-rays are due to an internal plateau, this requires that the FS emission from the jet must be fainter than the observed X-ray detections (Figure \ref{fig:xraylc}) as well as all upper limits at any wavelength. In a standard GRB interpretation the FS component must exist, but in this case it is likely extremely faint due either to a low density \citep[a ``naked'' burst, e.g.,][]{KumarPanaitescu2000,Perley2009}, low fraction of energy in magnetic fields $\varepsilon_B$ \citep[e.g.,][]{Barniol2014}, or an early jet break (which should also impact the emission from the long-lived engine). 

The lack of a known distance and the known degeneracy and broad allowed parameter range in afterglow modeling preclude a conclusive diagnostic of this possibility though experience has shown that it is generally not difficult to ``hide'' a forward shock, especially when applying a rarefied environment \citep[e.g.,][]{KumarPanaitescu2000,Perley2009}. In what follows we briefly test this possibility. 

\begin{figure*}
    \centering
\includegraphics[width=\columnwidth]{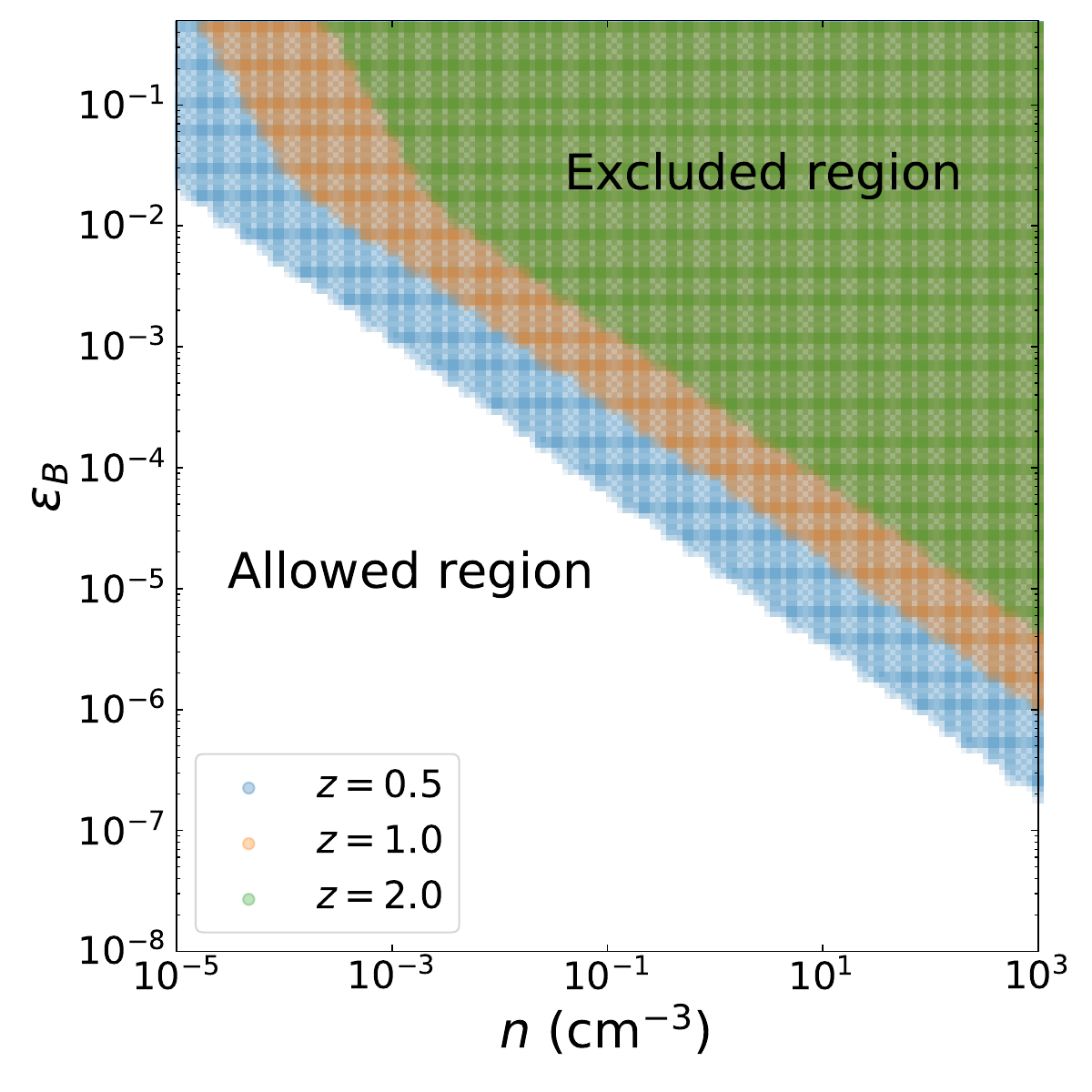} 
\includegraphics[width=\columnwidth]{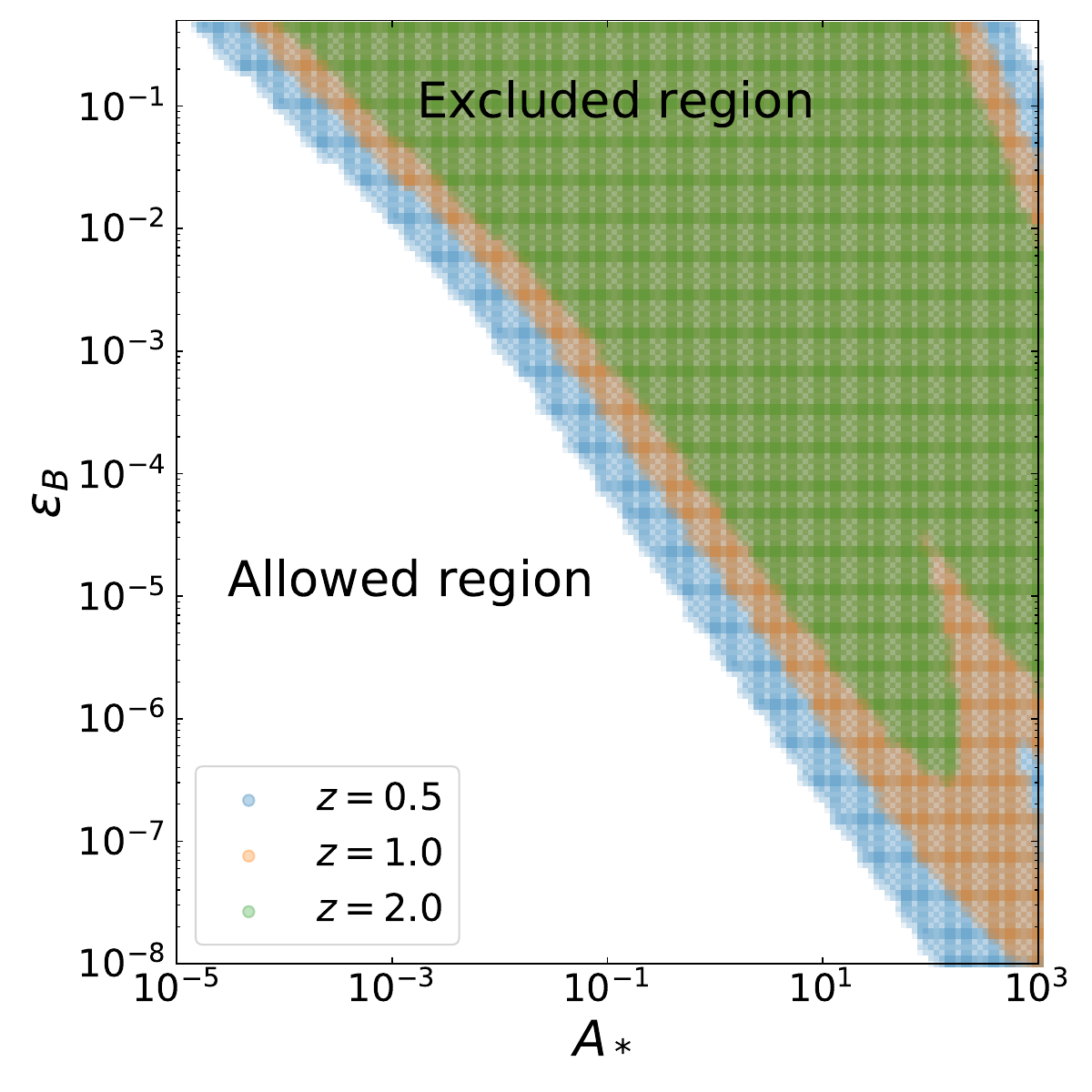}
    \caption{\textbf{Left:} Allowed parameter space for non-detection of an afterglow in a uniform density environment ($k$\,$=$\,$0$). Here we have fixed $E_\textrm{kin}$\,$=$\,$10^{52}$ erg, $p$\,$=$\,$2.2$, and $\varepsilon_\textrm{e}$\,$=$\,$0.1$. 
    \textbf{Right:} Same as the left figure but for a wind-like environment ($k$\,$=$\,$2$). 
    }
    \label{fig:allowedafterglow}
\end{figure*}

In the standard fireball model \citep[e.g.,][]{Meszaros1997,Wijers1999,Granot2002}, where an on-axis tophat jet propagates into an external medium $\rho_\textrm{ext}(R)$\,$=$\,$A\, R^{-k}$, 
the forward shock emission at each time and frequency is determined by a set of 6 parameters: $\{z,p,E_\textrm{kin},A,\varepsilon_\textrm{e},\varepsilon_\textrm{B}\}$, where $p$ is the slope of the electron's powerlaw energy distribution, $E_\textrm{kin}$ is the kinetic energy, and $\varepsilon_\textrm{e}$ and $\varepsilon_\textrm{B}$ represent the fraction of shock energy in electrons and magnetic fields, respectively. While the powerlaw index $k$ determining the structure of the surrounding environment is a free parameter for simplicity we adopt $k$\,$=$\,$0$ for a uniform density environment 
and $k$\,$=$\,$2$ for a wind-like environment 
\citep[e.g.,][]{ChevalierLi2000}. 
These are the standard limiting cases, and we apply the analytic results of \citet{Granot2002} in our analysis. We have assumed a post-deceleration and pre-jet-break behavior of the outflow. We neglect the effect of viewing angle, inverse Compton corrections, or reverse shock emission.

The observed synchrotron emission from the forward shock at a given time and frequency depends on the relation between the characteristic synchrotron frequencies: the self-absorption frequency $\nu_\textrm{a}$, the injection frequency $\nu_\textrm{m}$, and the cooling frequency $\nu_\textrm{c}$. We account for Spectra $1$\,$-$\,$5$ as outlined in Figure 1 of \citet{Granot2002}. We consider a range of redshifts $z$\,$=$\,$\{0.5,1.0,2.0\}$ and kinetic energies $E_\textrm{kin}$\,$=$\,$\{10^{51},10^{52},10^{53}\}$ erg, and fix $p$\,$=$\,$2.2$ and $\varepsilon_\textrm{e}$\,$=$\,$0.1$ for simplicity \citep{BeniaminiVanderHorst2017,Duncan2023}. We compute the afterglow flux at the time and frequency of our VLA and ATCA upper limits ($10$\,$-$\,$260$ d; observer frame) and include the X-ray limits from \textit{Swift} ($>$\,$10$ d). We also use the observed X-ray detections from \textit{Swift} ($1.4$ d) and \textit{NICER} as upper limits to the forward shock emission. While we do not account for the optical and near-infrared upper limits as they can be significantly impacted by intrinsic dust in the host galaxy we do find that generally these limits are automatically satisfied. We compute the detectability over a grid of 100 log-uniform steps between $n$\,$=$\,$\{10^{-5},1000\}$ cm$^{-3}$ for a uniform environment ($k$\,$=$\,$0$), $A_*$\,$=$\,$\{10^{-5},1000\}$ for a wind environment ($k$\,$=$\,$2$), and $\varepsilon_\textrm{B}$\,$=$\,$\{10^{-8},0.5\}$ for both values of $k$.

We emphasize that these calculations are relevant to both GRB and TDE interpretations for EP240408a (see also \S \ref{sec:tde}). 
In general the most constraining limit is from the VLA at 10.34 d (observer frame) post-trigger, and the late-time VLA observations at 158 d and 258 d (observer frame) only excludes a small parameter space (for a wind environment) at high densities where $\nu_\textrm{a}$ was $\sim$\,$100$ GHz at early times. The late-time VLA upper limits are more constraining in a TDE interpretation (see \S \ref{sec:tde}), as most GRBs experience a jet break before $158$\,$-$\,$258$ d and we have not accounted for post-jet-break behavior in our calculations which would lead to dim radio emission at late times.

We find that for a uniform environment the forward shock emission is generally in the regime $\nu_\textrm{a}$\,$<$\,$\nu_\textrm{m}$\,$<$\,$\nu_\textrm{c}$ (Spectrum 1 from \citealt{Granot2002}) whereas for a wind environment solutions are also found for $\nu_\textrm{m}$\,$<$\,$\nu_\textrm{a}$\,$<$\,$\nu_\textrm{c}$ (Spectrum 2 from \citealt{Granot2002}). In either case (see Figure \ref{fig:allowedafterglow}) the solutions require either very low density environments or low values of $\varepsilon_\textrm{B}$. We note that values of $\varepsilon_\textrm{B}$ in the range $10^{-1}$ to $10^{-4}$ are generally considered standard, but have been found to be as low as $10^{-8}$ in some GRBs \citep{BarniolDuran2013,Santana2014,Beniamini2015,Zhang2015}.

The solutions can span a larger amount of  parameter space if $\varepsilon_\textrm{e}$ and $p$ are allowed to vary over a larger range of values. In general these calculations show that we cannot completely exclude an on-axis forward shock emission from a post-deceleration jet. The allowed parameter space would be greatly expanded by allowing for off-axis viewing angles.

\subsubsection{An Off-axis Jet}

\begin{figure}
    \centering
\includegraphics[width=1\columnwidth]{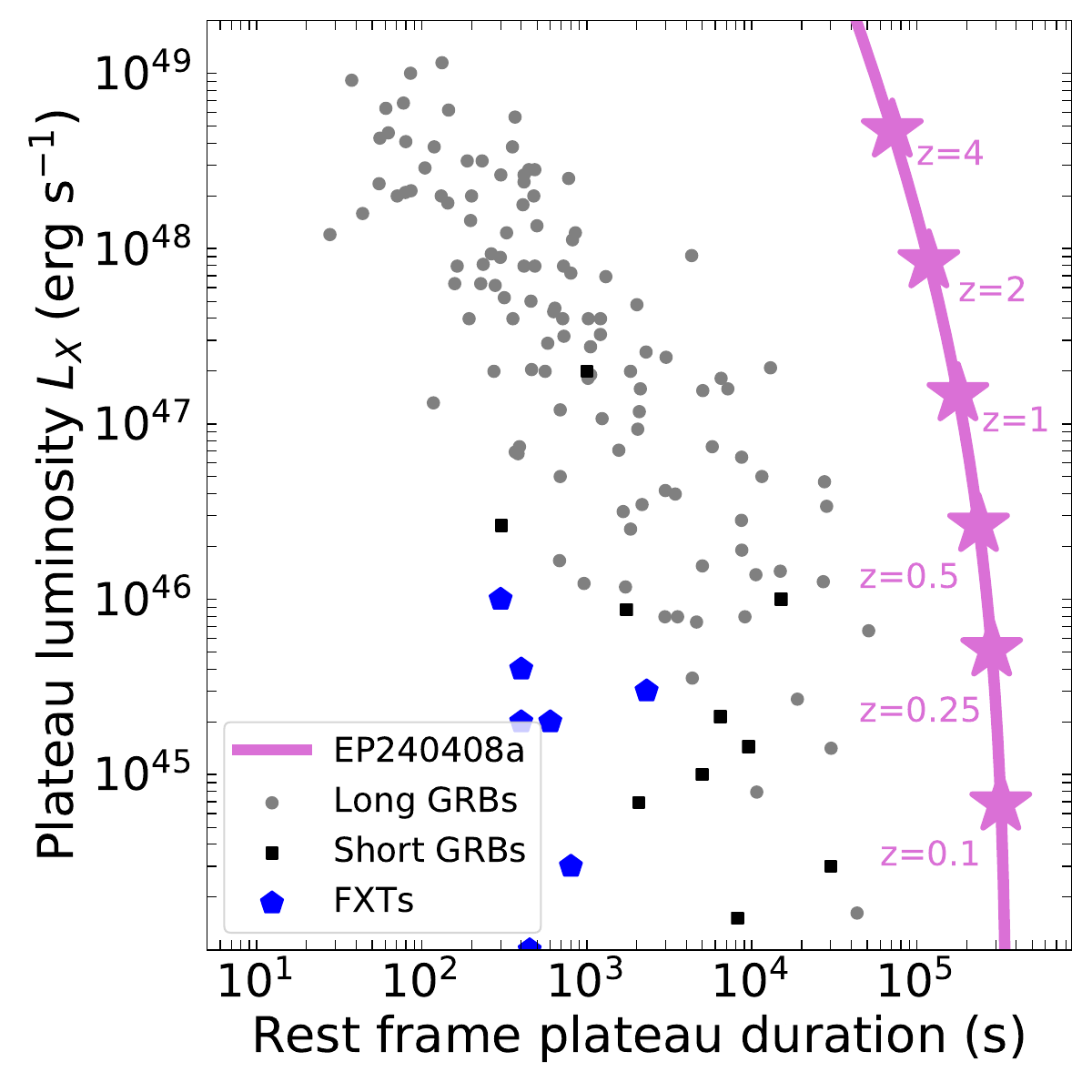}
    \caption{Observed X-ray plateau rest frame duration and luminosity for GRBs \citep{Tang2019,Xu2021} and FXTs  \citep{Quirola2024} versus EP240408a at multiple redshifts. EP240408a does not follow the standard plateau correlations of either short or long GRBs. We note that $z$\,$\lesssim$\,$0.5$ is strongly disfavored for EP240408a. 
    }
    \label{fig:grbplat}
\end{figure}

In Figure \ref{fig:grbplat} we compare the rest frame duration and luminosity of X-ray plateaus observed from both short and long GRBs to EP240408a. EP240408a is an outlier and would have the longest plateau ever discovered from a GRB, even longer than the class of ultralong GRBs (e.g., GRBs 060729, 101225A, 121027A, 130925A, 141121A), which are though to have the longest lasting central engines \citep[e.g.,][]{Levan2014,Cucchiara2015}. However, the major difference from ultralong GRBs is the lack of energetic $\approx$\,$10^{54-55}$ erg (isotropic-equivalent) gamma-ray emission. 

The lack of gamma-rays is hard, though not impossible, to explain in the context of a GRB. For example, an off-axis viewing angle would decrease the observed, line-of-sight (isotropic-equivalent) energy of the GRB, potentially allowing for consistency with the upper limits in Figure \ref{fig:grb}, which are compared to a sample of on-axis bursts. An additional, and quite relevant, impact of viewing a GRB off-axis is a likely decrease in the observed peak energy $E_\textrm{p}$ towards soft X-rays (assuming $E_\textrm{p}$ is constant in the comoving frame; see, e.g., \citealt{Parsotan2024} for a discussion of the results of numerical simulations, and \citealt{Fan2023,Gao2024} for discussions on the observational impact), though this will depend on the exact Lorentz factor profile of the jet. The obvious alternative is a low gamma-ray efficiency, which could be either an impact of an off-axis viewing angle \citep{BeniaminiNakar2019,BBPG2020,Gill2020,OConnor2024} or due to inefficient breakout of the jet from the progenitor star(s). 

The tension between the length of the X-ray plateau of EP240408a and those of the larger sample (Figure \ref{fig:grbplat}) can also be reduced if EP240408a is seen off-axis. An off-axis viewing angle delays the arrival of radiation to the observer and can extend this phase \citep[e.g.,][]{Nakar2002,Panaitescu2003,vanEerten2010}. 
This would also produce less luminous X-ray emission compared to the on-axis cases shown in Figure \ref{fig:grb} (right panel), which would 
likely require a very nearby event as far off-axis angles are strongly disfavored for the observed cosmological GRB sample \citep[e.g.,][]{Ryan15,Matsumoto2019a,OConnor2024}. 

In addition, the lack of a likely host galaxy, or even any potential low redshift host, excludes a low luminosity GRB such as GRB 060218 \citep{Soderberg2006grb060218} which at $z$\,$=$\,$0.033$ had a gamma-ray energy of $6\times10^{49}$ erg and 11 hour X-ray luminosity of $L_\textrm{X,11}$\,$\approx$\,$10^{42}$ erg s$^{-1}$. At a similar redshift EP204048a would have a gamma-ray energy $<$\,$10^{48-49}$ erg and an X-ray luminosity at 11 hr of $\sim$\,$10^{44}$ erg, quite different from even the lowest luminosity GRBs. Such low redshifts (even up to $z$\,$<$\,$0.5$) are also excluded by the lack of detection of a coincident supernova (Figure \ref{fig:ULs}; right panel) in our deep multi-epoch Gemini imaging which probes the typical peak times of known GRB-SN \citep[e.g.,][]{Galama1998,Soderberg2006grb060218,Hjorth2012sn,Cano2017,Srinivasaragavan2023,Srinivasaragavan2024}. 

\subsubsection{A Long-lived Magnetar}
\label{sec:magnetar}

The most commonly adopted mechanism to explain the long-lived engine required by internal plateaus is a newborn, rapidly spinning ($P$\,$\approx$\,$1$\,$-$\,$10$ ms) magnetar \citep{Usov1992,Duncan1992} with a high magnetic field $B$\,$\approx$\,$10^{15-16}$ G \citep{Zhang2006,Liang2006,Troja2007,Lyons2010,Metzger2011,Rowlinson2010,Rowlinson2013,Giacomazzo2013,Lu2015,Chen2017}. In many cases the required magnetar approaches the maximum theoretical limits, and other interpretations for internal plateaus have been suggested \citep[e.g., black hole powered photospheric emission or low Lorentz factor leading to a late deceleration of the jet;][]{Shen2012,Duffell2015,Beniamini2017,BeniaminiGiannios2017,Beniamini2019plateau,Oganesyan2020,Dereli-Begue2022}.

In this case of a magnetar, the plateau is produced by spin-down of the magnetar's dipole field \citep{Zhang2001}
\begin{align}
    L_\textrm{sd} = 1.0\times10^{49}\, B^2_{15} P_{-3}^{-4} R_6^6\;\textrm{erg}\,\textrm{s}^{-1},
\end{align}
where $R$ is the radius of the neutron star. 
We have adopted the convention $B_{15}$\,$=$\,$B/(10^{15}\,\textrm{G})$, and similar for the other parameters using cgs units. 
We note that the efficiency of converting the spin-down power to the observed X-ray luminosity $L_\textrm{X}$\,$=$\,$\eta L_\textrm{sd}$ is generally a challenge \citep{Granot2015,Beniamini2017}. 
The end of the plateau phase is typically taken to be the spin-down timescale of the magnetar \citep{Zhang2001}
\begin{align}
    t_\textrm{sd} = 2.0\times10^{3}\,I_{45}\,B_{15}^{-2}P_{-3}^2R_6^{-6}\;\textrm{s}.
\end{align}
where $I$ is the moment of inertia of the neutron star. 
The temporal decay following the spin-down time is \citep{Beniamini2017}
\begin{align}
    L=L_\textrm{sd}\Bigg(1+\frac{n-1}{2}\frac{t-t_\textrm{o}}{t_\textrm{sd}}\Bigg)^{\frac{1+n}{1-n}}
\end{align}
where $t_\textrm{o}$ is the start of the spin-down, and $n$ is the magnetic braking index which for $n$\,$=$\,$3$ yields the standard $\propto$\,$t^{-2}$ decay. A $t^{-7}$ decay as observed for EP240408a requires $n$\,$\approx$\,$1.33$, consistent with that inferred for some isolate pulsars \citep{Hamil2015}. In addition, the temporal decline can steepen if the magnetar collapses to a black hole, but this would limit the allowed energy reservoir, which considering the estimate $P$\,$\sim$\,$1$\,ms is already constraining.

Despite the long-lasting plateau ($\sim$\,$3.5\times10^5$ s), we find a suitable match to the required luminosity and duration for a magnetar with $B$\,$\approx$\,$10^{14}$ G and $P$\,$\approx$\,$1$ ms assuming $z$\,$\approx$\,$1$. 
These solutions are degenerate with redshift and radiative efficiency, and thus we only provide a single example. In any case, the initial spin period of the neutron star would have to be close to the breakup limit \citep{LattimerPrakash2004}. These values, in particular the magnetic field, are slight outliers when compared to those inferred for either short  \citep[$B$\,$\approx$\,$10^{16}$ G;][]{Rowlinson2010,Rowlinson2013,Gompertz2014,Lu2015} or long GRBs \citep[$B$\,$\approx$\,$10^{15}$ G;][]{Lyons2010,Yi2014} and fall closer to SLSN \citep[$B$\,$\approx$\,$10^{13-14}$ G;][]{Nicholl2017}. This is not surprising given the longer plateau duration $t_\textrm{sd}$\,$\propto$\,$B^{-2}$ which requires a smaller magnetic field.

\subsubsection{A Dissipative Photosphere}
\label{sec:photospheregrb}

There are other possible explanations for the observed plateau other than a long-lived magnetar or an off-axis jet.
A black hole engine was proposed for GRB 070110 \citep{Beniamini2017} whereas the central engine launches a second lower Lorentz factor outflow which produces photospheric emission \citep{BeniaminiKumar2016}, leading to the observed plateau. We note however that this model is independent of the class of central engine, which simply provides an energy source that must then be transformed (e.g., a dissipative photosphere) into the observed radiation. 
\citet{Beniamini2017} considered a jet which efficiently dissipates energy below the photosphere leading to a thermal Comptonized X-ray spectrum with peak energy $E_\textrm{p,X}$\,$\lesssim$\,$0.5$\,$-$\,$1$ keV. In the observed $0.3$\,$-$\,$10$ keV bandpass such a spectrum would appear as $F_\nu$\,$\approx$\,$\nu^{-1}$ (similar to EP240408a). However, in the case of EP240408a, the peak energy is $\sim$\,$4$ keV (Figure \ref{fig:xrayspectra}).

Applying the model from \citet{Beniamini2017}, we can derive the Lorentz factor $\Gamma$ at the photosphere required to produce the observed X-ray plateau luminosity $L_\textrm{th}$ as 
\begin{align}
\label{eqn:photgamma}
    \Gamma = 15 \frac{\lambda^{1/2}}{\epsilon_\textrm{rad,-2}^{1/4}}(1+\sigma)^{-1/4}L_\textrm{th,47}^{1/8}E_\textrm{p,X}^{1/2}
\end{align}
where $\lambda$ is an order unity factor determined by the exact emission spectrum\footnote{The parameters $\lambda$ is $1$ for a pure non-Comptonized photosphere and less than 1 for a Comptonized photosphere.}, $\epsilon_\textrm{rad}$ is approximately the efficiency of a non-dissipative photosphere, and $(1+\sigma)$ is the magnetization parameter. In this model, we find that the observations of EP240408a could be roughly reproduced by a photosphere with Lorentz factor $\Gamma$\,$\approx$\,$30$.

As discussed in \citet{Beniamini2017} the Lorentz factor is not severely impacted by the allowed ranges of $\lambda$, $\epsilon_\textrm{rad}$, or $(1+\sigma)$.  These parameters instead have a larger impact on the (observer frame) geometric timescale \citep{Beniamini2017}
\begin{align}
\label{eqn:tgeo}
    t_\textrm{geo} = 0.25 \frac{1+z}{2} \frac{\epsilon_\textrm{rad,-2}(1+\sigma)^{1/4}}{\lambda^{5/2}}\frac{L_\textrm{th,47}^{3/8}}{E_\textrm{p,X}^{5/2}}\;\textrm{s}
\end{align}
However, the geometric timescale is always significantly shorter than the plateaus decay timescale, which strongly suggests that the rapid $t^{-7}$ decay is due to the cessation of the central  engine.

\subsection{A High Redshift Jetted Tidal Disruption Event}
\label{sec:tde}

The population of relativistic jetted TDEs is  small, with only four candidates \citep{Bloom2011,Levan2011,Zauderer2011,Burrows2011,Cenko2012,Brown2015,Pasham2015,Andreoni2022,Pasham2023} uncovered since their initial discovery in 2011 \citep{Bloom2011,Levan2011,Burrows2011}. They are produced by the tidal shredding of a stellar mass star by a massive black hole ($M_\textrm{BH}$\,$\approx$\,$10^{5-8} M_\odot$). 
These TDEs are generally characterized by luminous X-ray and radio emission (Figures \ref{fig:xraylc} and \ref{fig:ULs}). At X-ray wavelengths they display rapid short term variability ($\sim$ hours) on top of a powerlaw decay of $\approx$\,$t^{-2}$.

Only Sw J1644+57 ($z$\,$=$\,$0.35$) was detected by a gamma-ray satellite (\textit{Swift}/BAT; \citealt{Cummings2011J1644}) in flight, as opposed to a ground analysis. Even then, Sw J1644+57 was identified through a $\sim$\,$1000$ s long image trigger \citep{Cummings2011J1644,Sakamoto2011J1644,Levan2011} and not bright enough to trigger \textit{Swift}/BAT as a normal GRB. Accompanying its multiple gamma-ray triggers, Sw J1644+57 exhibited exceptional soft X-ray ($0.3$\,$-$\,$10$ keV) radiation with peak luminosity $L_\textrm{X}$\,$\approx$\,$3\times10^{48}$ erg s$^{-1}$ and an average luminosity over $10^6$ s of $9\times10^{46}$ erg s$^{-1}$ \citep{Bloom2011}. We note that the observed peak X-ray flux of the initial trigger of EP240408a is a factor of $\sim$\,$10$ higher than the brightest soft X-ray detection of Sw J1644+57. 

Two additional jetted TDEs, Sw J2058+05 at $z$\,$=$\,$1.19$ \citep{Cenko2012} and Sw J1112-82 at $z$\,$=$\,$0.89$ \citep{Brown2015}, were both identified in an automated ground based analysis by the \textit{Swift}/BAT Hard X-ray Transient Monitor \citep{Krimm2013} in 4 day binned windows ($15$\,$-$\,$50$ keV). As such their gamma-ray variability or spectra could not be measured, and they cannot be obviously compared to the population of GRBs in Figure \ref{fig:grb}.

In contrast to the other three jetted TDEs, AT2022cmc was discovered as a rapidly fading optical transient and was not detected in gamma-rays \citep[$z$\,$=$\,$1.19$;][]{Andreoni2022,Pasham2023,Rhodes2023,Yao2024}. \citet{Andreoni2022} estimated a $<$\,$5\%$ chance of \textit{Swift}/BAT detecting similar gamma-ray flares to those observed from Sw J1644+57 at redshift $z$\,$=$\,$1.19$. Due to its larger distance AT2022cmc remains hostless to depth 24.5 AB mag \citep{Andreoni2022}. In the case of AT2022cmc, despite the subarcsecond localization, the  lack of an underlying host precluded the determination as to whether it was truly a nuclear transient. In the case of EP240408a, the lack of subarcsecond localization (Figure \ref{fig:fc}) is the limiting factor, and the slight offset from the XRT enhanced position similarly does not rule out EP240408a as a nuclear transient. 

Therefore, we find that the lack of gamma-rays does not immediately disfavor a jetted TDE interpretation, nor does the prompt soft X-ray detection. 
If indeed the X-rays can shutoff on timescales of $\sim$\,$4$ d there is a strong bias against finding such fast X-ray transients. 
In fact, there are selection biases that exist, especially prior to the launch of \textit{Einstein Probe} \citep{EP2015,EP2022}, towards detecting similar prompt X-ray transients to EP240408a \citep{GCN36053eptrigger} or relativistic jetted TDEs in general.

\subsubsection{The Nature of the X-ray Emission}

While the X-ray radiation observed from EP240408a can match the luminosity of known relativistic jetted TDEs (Figure \ref{fig:xraylc}) at $z$\,$\approx$\,$1$\,$-$\,$2$, the observed timescales are significantly shorter (Figure \ref{fig:xraylc}). While after $\sim$\,$10$ d, Sw J1644+57 transitioned to an approximately $t^{-5/3}$ decay \citep{Burrows2011,Bloom2011}, EP240408a appears to potentially shut off with an extremely fast decay ($\delta \tau/\tau$\,$\approx$\,$0.2$). In particular, central engine cessation of other jetted TDEs is on the order of hundreds of days \citep[$100$\,$-$\,$400$ d in the rest frame;][]{mangano2016,Levan2016,Eftekhari2018,Eftekhari2024}, whereas a sharp decline is observed in EP240408a after only $\lesssim$\,$4$ d when considering the impact of redshift. 
It is unclear whether this is natural variability in X-rays, like the sharp variability observed at early times for Sw J1644+57 (Figure \ref{fig:xraylc}), at high-$z$ that then fades below \textit{NICER}, \textit{Swift}, and \textit{NuSTAR} sensitivity. 

For starters, if we assume a typical black hole mass for the observed relativistic jetted TDEs $M_\textrm{BH}$\,$\approx$\,$10^{6-8} M_\odot$ the Eddington luminosity is in the range $10^{44-46}$ erg s$^{-1}$. Even ignoring the initial EP trigger, the X-ray plateau luminosity implies a highly super-Eddington outflow with Eddington ratio $\lambda$\,$\approx$\,$5$\,$-$\,$500$ ($20$\,$-$\,$2,000$) at $z$\,$\approx$\,$1.0$ ($2.0$). This may favor redshifts more similar to AT2022cmc at $z$\,$=$\,$1.19$ as the Eddington ratio at $z$\,$\gtrsim$\,$2$ is quite extreme. 
We note that the initial EP trigger is $\sim$\,$500$ times brighter and would require extreme super-Eddington accretion. For reference, Sw J1644+57 exhibited an  Eddington ratio (beaming corrected) of only $\sim$\,$1,200$ \citep{Beniamini2023TDE}.

If we assume the emission comes from a highly collimated, relativistic jet (similar to Sw J1644+57), it significantly decreases the required energy budget. For instance, the beaming factor of a tophat jet is $f_b$\,$\approx$\,$\theta_\textrm{c}^2/2$, where $\theta_\textrm{c}$ is the jet's core half-opening angle. Adopting a typical opening angle of $\theta_\textrm{c}$\,$=$\,$0.1$ rad yields a beaming correction of $f_b^{-1}$\,$\approx$\,$200$. Applying this to the initial flare detection of EP240408a leads to an Eddington ratio of between $\lambda_\textrm{peak}$\,$\approx$\,$13$\,$-$\,$1,250$ ($50$\,$-$\,$5,000$) at $z$\,$\approx$\,$1.0$ ($2.0$). 
These ratios can be decreased further if the opening angle of the jet is smaller than $0.1$ rad ($6$ deg). 
This implies the observed X-ray emission requires a relativistic jet (with likely a small viewing angle) and that we are observing X-rays from an internal dissipation process (due to the short variability timescales observed for the plateau).

\subsubsection{Fallback Timescale}
\label{sec:fb}

The end of the plateau can be associated to the fallback timescale of the disrupted stellar material \citep[e.g.,][]{Burrows2011,Bloom2011,Cenko2012} from a main sequence star in which case the fallback accretion rate begins to decay as $\approx$\,$t^{-\alpha}$ where $\alpha$\,$=$\,$5/3$ (complete disruption) or $2.2$ (partial disruption) is generally adopted \citep{Guillochon2013}. 
The fallback time is given by \citep{Ulmer1999,Stone2013,Stone2016}
\begin{align}
    t_\textrm{fb} = 3.5\times10^{5}\; \textrm{s}\; \Big(\frac{M_\textrm{BH}}{10^4 M_\odot}\Big)^{1/2} \Big(\frac{M_*}{ M_\odot}\Big)^{-1} \Big(\frac{r_*}{ R_\odot}\Big)^{3/2} 
\end{align}
where we can further adopt $r_*$\,$=$\,$M_*^{0.8}$ for a main sequence star \citep{Stone2013,Stone2016}.
Thus, either we require an intermediate mass black hole or a sub-solar mass star. The disruption of a sub-solar mass star is plausible based on the standard initial mass function \citep{Chabrier2003}. The Eddington ratio scales as $\lambda$\,$\propto$\,$M_\textrm{BH}^{-3/2}$ and thus smaller black holes are capable of higher accretion rates leading to greater super-Eddington luminosities. 

An alternative possibility is the disruption of a white dwarf by a black hole \citep[e.g.,][]{Ye2023}
\begin{align}
\label{eqn:fbwd}
    t_\textrm{fb} = 1.0\times10^{5}\; \textrm{s}\; \Big(\frac{M_\textrm{BH}}{10^7 M_\odot}\Big)^{1/2} \Big(\frac{M_\textrm{WD}}{0.6 M_\odot}\Big)^{-1} \Big(\frac{r_\textrm{WD}}{10^9\,\textrm{cm}}\Big)^{3/2} 
\end{align}
where we apply the white dwarf mass-radius relation following \citet{Nauenberg1972}.  In this case, in order to achieve a late enough fallback timescale requires a larger black hole. A similar value of $10^{5}$ s is derived for $M_\textrm{BH}$\,$\approx$\,$10^{6} M_\odot$ and $M_\textrm{WD}$\,$\approx$\,$0.3 M_\odot$. 

\subsubsection{Transition from Super-Eddington to Sub-Eddington Accretion}
\label{sec:edd}

An alternative is the cessation of the central engine, which can be interpreted at the transition from super-Eddington accretion to sub-Eddington accretion \citep[e.g.,][]{Zauderer2013,Pasham2015,Eftekhari2024}. This can be taken as the shutoff of a relativistic jet. This occurs at an Eddington ratio $\lambda$ of unity ($\lambda$\,$=$\,$1$), such that 
\citep{Eftekhari2024} 
\begin{align}
\label{eqn:mbhmb}
    M_\textrm{BH} = 8.4\times10^{6}\, L_\textrm{jet,off,47}\Big(\frac{\varepsilon_\textrm{disk,-1}}{\varepsilon_\textrm{jet,-1}}\Big)\Big(\frac{f_\textrm{b}}{200}\Big)f_\textrm{bol,3} \; M_\odot
\end{align}
where $\varepsilon_\textrm{disk}$ and $\varepsilon_\textrm{jet}$ represent the radiative efficiencies of the disk and jet, respectively. \citet{Eftekhari2024} finds that the $\varepsilon_\textrm{jet}/\varepsilon_\textrm{disk}$\,$\gtrsim$\,$0.1$. We convert the isotropic equivalent luminosity of the jet to the intrinsic luminosity using a tophat jet beaming factor $f_\textrm{b}$\,$=$\,$(1-\cos\theta_\textrm{c})$\,$\approx$\,$\theta_\textrm{c}^2/2$. As before we adopt a jet half-opening angle $\theta_\textrm{c}$\,$=$\,$0.1$ rad such that $f_\textrm{b}^{-1}$\,$=$\,$200$. An additional bolometric correction to the jet energy is $f_\textrm{bol}$\,$=$\,$3f_\textrm{bol,3}$ is also made \citep{Eftekhari2024}.

Following \citet{Eftekhari2024}, the time of the transition to sub-Eddington accretion occurs at 
\begin{align}
    t_\textrm{off} = t_\textrm{fb} \lambda_\textrm{p}^{1/\alpha}
\end{align}
where $\lambda_\textrm{p}$ is the peak Eddington ratio \citep{Stone2013,Stone2016}
\begin{align}
    \lambda_\textrm{p} = 133\, f_\textrm{in} \varepsilon_\textrm{disk,-1} M_\textrm{BH,6}^{-3/2}  \Big(\frac{M_*}{ M_\odot}\Big)^{2}\Big[\frac{3(\alpha-1)}{2}\Big]
\end{align}
where $f_\textrm{in}$ accounts for multiple efficiencies related to the fallback process (see \citealt{Eftekhari2024} for further discussion). 

This interpretation immediately requires a short fallback time, which can potentially be accommodated by the disruption of a white dwarf by an intermediate mass black hole (Equation \ref{eqn:fbwd}). For consistency with Equation \ref{eqn:mbhmb}, this in turn likely also requires either a smaller redshift or a significantly narrower jet, both of which lead to a lower jet luminosity and smaller required black hole mass. Assuming the disruption of a $0.6M_\odot$ white dwarf by an intermediate mass black hole $\approx$\,$10^{5} M_\odot$, as favored by \citet{Eftekhari2024} to explain other jetted TDEs, yields $t_\textrm{fb}$\,$\approx$\,$10^4$ s and a shutoff time of $t_\textrm{off}$\,$\approx$\,$2.7\times10^{5}$ s for $\alpha$\,$=$\,$2.2$. Given the multiple uncertainties related to the efficiency of the fallback process it is possible to shorten this timescale further, allowing it to match the rest frame shutoff time, $3.5\times10^{5}/(1+z)$ s, of EP240408a over a variety of redshifts. 

One possibility to reduce these constraints is that the initial EP trigger is related to the peak of the fallback accretion rate and that the initial disruption of the star occurred on earlier timescales. However, an earlier disruption time immediately implies that the temporal slopes determined in this work are shallower than reality, and that the break decay must be steeper than $t^{-7}$, which is already challenging to explain. We note that this can allow for the initial slope, prior to the steep decline, to match either $t^{-5/3}$ or $t^{-2.2}$ (for explosion times between $\sim$\,$3$\,$-$\,$10$ d prior to the EP trigger). The major issue with this interpretation lies in explaining the factor of $\sim$\,$1,000\times$ flare of the EP trigger if it occurs a few days after disruption. Such a flare has not been previously observed in any other jetted TDE.

An alternative scenario proposed by \citet{Teboul2023} suggests that early precession of the jet may have delayed its breakout from the surrounding ejecta. The breakout time for a precessing jet can be on the order of a few days to tens of days \citep{Teboul2023},  increasing the jet shutoff time to more reasonable values, and decreasing the strong constraints on the progenitor system (e.g., black hole mass). This model also explains the lack of early short term variability, suggesting that the rapid X-ray brightness variations from Sw J1644+57 were due to an early jet breakout while the jet was still precessing across the line-of-sight (Figure \ref{fig:xraylc}), whereas jet breakout after alignment with the black hole's spin leads to a smooth decay of the X-ray lightcurve as observed from AT2022cmc and EP240408a.

\begin{figure*}
    \centering
\includegraphics[width=\columnwidth]{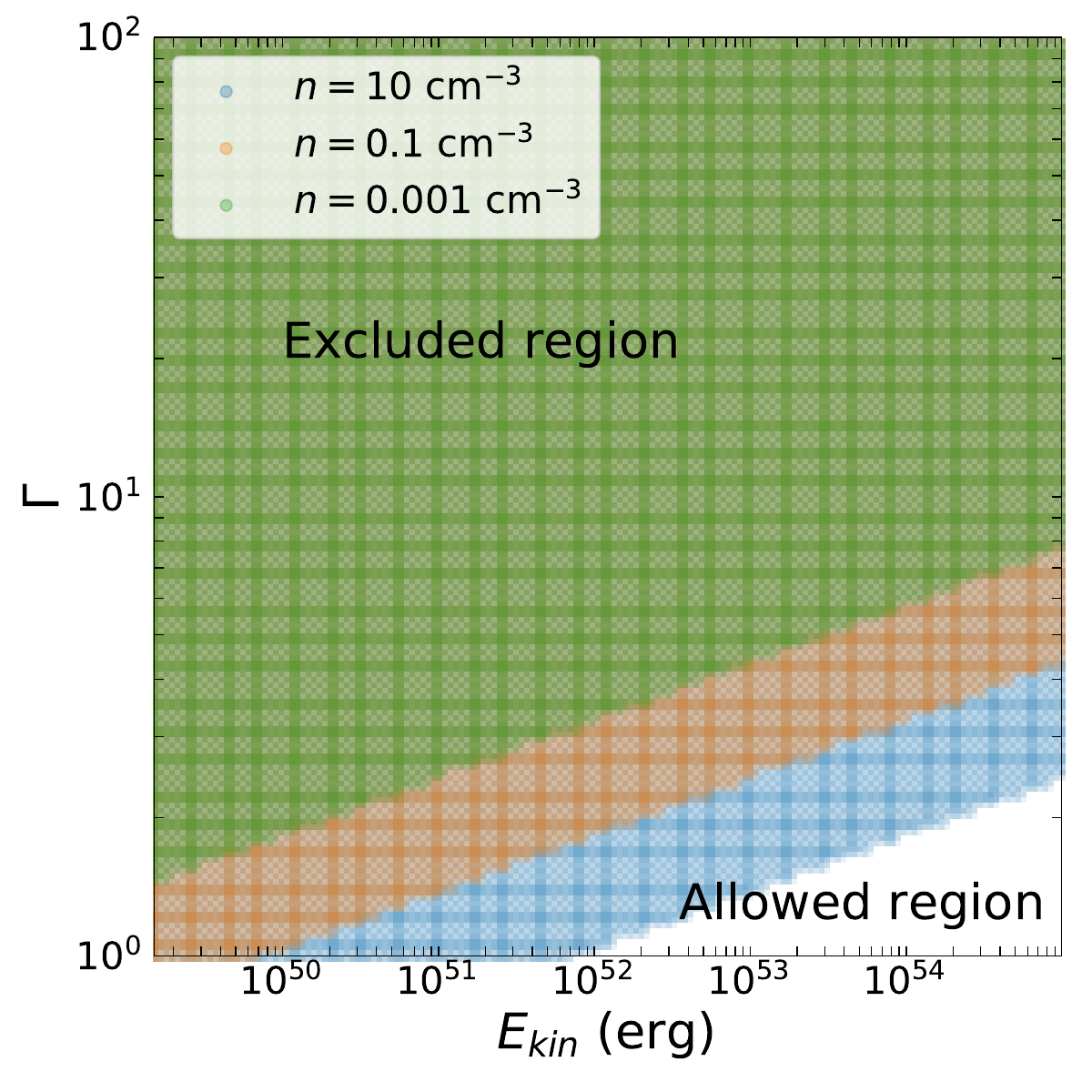} 
\includegraphics[width=\columnwidth]{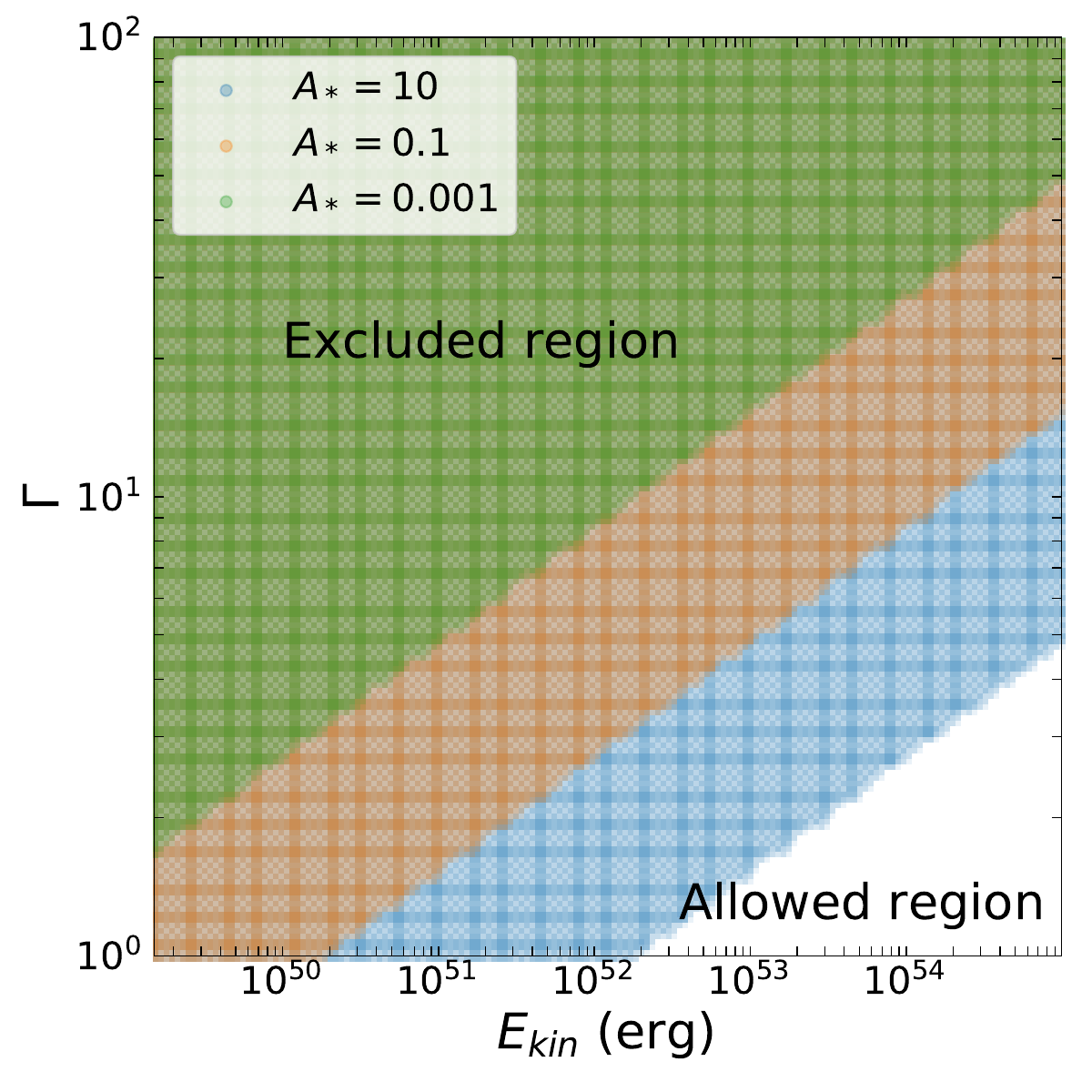}
    \caption{\textbf{Left:} Allowed parameter space in Lorentz factor $\Gamma$ and isotropic-equivalent kinetic energy for delayed jet deceleration of an ultra relativistic outflow to after $>$\,$158$\,$-$\,$258$ d (observer frame) in a uniform density environment ($k$\,$=$\,$0$). Here we have fixed $z$\,$=$\,$1$ due to the larger dependence and allowed range of the other parameters. 
    \textbf{Right:} Same as the left figure but for a wind-like environment ($k$\,$=$\,$2$). 
    }
    \label{fig:alloweddecel}
\end{figure*}

\subsubsection{Delayed Radio Emission}

Due to the lack of radio emission detected for EP240408a (Figure \ref{fig:ULs}) and the short variability timescales ($\delta \tau/\tau$\,$\approx$\,$0.2$), we interpret the X-rays as arising from an internal process within the jet. 
The lack of optical emission (Figure \ref{fig:ULs}) can be due to intrinsic extinction in the host galaxy, as was the case for Sw J1644+57. The expected $A_V$ from the inferred X-ray hydrogen column density (\S \ref{sec:xrayspectra}) is $1.4$ ($6.3$) mag \citep{Guver2009} at $z$\,$\approx$\,$0.5$ ($2.0)$, and can easily lead to a non-detection at higher redshifts. 
Harder to explain is the lack of bright radio in our sensitive multi-epoch VLA and ATCA observations. It is possible that these observations occur too early (i.e., pre-deceleration) and that eventually rising radio emission could become detectable, or that the self-absorption frequency is above our observations at $5$\,$-$\,$18$ GHz out to $\sim$\,$260$ d (observer frame).

Delayed radio emission has been uncovered in a variety of TDEs \citep{Horesh2021a,Horesh2021b,Perlman2022,Sfaradi2022,Cendes2022,Cendes2024} on timescales of $>$\,$100$\,$-$\,$1,000$ d. This delayed radio emission has been interpreted either as either late-jet launching from a long-lived central engine \citep{VanVelzen2016,Horesh2021a,Cendes2022,Cendes2024}, an off-axis relativistic jet with Lorentz factor on the order of tens \citep{Matsumoto2023,Beniamini2023TDE,Sfaradi2024}, or delayed breakout due early jet precession \citep{Teboul2023}. 
The further off-axis the viewing angle the later the time of the peak and the steeper the temporal slope of the rising emission \citep[e.g., AT2018hyz with a $t^{6}$ rise around 1,000 d after discovery;][]{Cendes2022,Sfaradi2024}. Due to the inferred range of Eddington ratios, it is likely that the jet of EP240408a cannot be very far off-axis. In addition, the alternative explanation that is typically invoked requires an outflow launched at late times \citep[e.g.,][]{Cendes2022,Cendes2024}, which conflicts with the already existing jet inferred from the X-rays.

As the jet is unlikely to be very far off-axis, we conclude that this is an unlikely reason for delaying the radio emission and that instead plausible interpretations include synchrotron self-absorption or late jet deceleration due to a low initial Lorentz factor $\Gamma$\,$\lesssim$\,$10$. 
We note that an equipartition analysis \citep{BarniolDuran2013,Barniol2013A,Matsumoto2023} is not  constraining as we have no robust limit on either the peak frequency  $\nu_\textrm{peak}$ or the peak flux $F_{\nu,\textrm{peak}}$. Instead, we simply compute the synchrotron emission from a forward shock propagating into an external medium $\rho_\textrm{ext}(R)$\,$=$\,$A (R/R_0)^{-k}$ for a range of parameters \citep{Granot2002}. We focus for simplicity on the limiting cases of $k$\,$=$\,$0$ for a uniform density environment and $k$\,$=$\,$2$ for a wind-like environment, though the exact index does not have to be either. The non-detection of forward shock emission requires low external densities or small fractions of energy in magnetic fields, as previously noted in \S \ref{sec:grb} and Figure \ref{fig:allowedafterglow}. These solutions are not impossible for a TDE jet and we therefore cannot rule out a post-deceleration forward shock.  An initially high synchrotron self-absorption ($>$\,$100$ GHz) frequency allows for non-detection of radio emission in a wind environment for large densities (Figure \ref{fig:allowedafterglow}; right panel), but requires a significant amount of dust to not be detected in the optical. This solution would predict luminous radio emission at higher frequencies or at the same frequencies at later times and is mainly ruled out by the late-time VLA upper limits (158 d and 258 d; observer frame). This high density solution disappears for higher kinetic energies (e.g., $10^{53}$ erg) or lower redshifts (e.g., $z$\,$<$\,$0.5$). Therefore, only for a wind-like environment with high densities $A_*$\,$>$\,$100$ and $E_\textrm{kin}$\,$<$\,$10^{53}$ erg do we predict detectable radio emission at later times or higher frequencies.

While these calculations (Figure \ref{fig:allowedafterglow}) assume that the jet is already in a post-deceleration phase, this may not be the case for a TDE jet. Thus we also consider the possibility that the jet has a low Lorentz factor and has not decelerated until $>$\,$260$ d (observer frame). In this case, the peak of the afterglow has not occurred and the emission is rising with time. This analysis mainly refers to a TDE interpretation, but could be relevant to GRBs in relation to low Lorentz factor dirty fireballs. We consider here the case of an on-axis jet, but note that off-axis viewing angles will delay the observed deceleration time further \citep[e.g.,][]{Nakar2002,OConnor2024}. 
The (on-axis) deceleration time (observer frame) of an ultra relativistic outflow with initial Lorentz factor $\Gamma$ is given by \citep{Sari1999,Molinari2007,Ghisellini2010,Ghirlanda2012,Nava2013,Nappo2014,Ghirlanda2018} 
\begin{align}
\label{eqn:tdec}
    t_\textrm{dec} = \left\{ \begin{array}{ll}  6.4\,\Big(\frac{1+z}{2}\Big) \Gamma_{1}^{-8/3}\,E_{\textrm{kin},53}^{1/3}\,n_{-1}^{-1/3} \; \rm{days}, & k=0, \\
15.2\,\Big(\frac{1+z}{2}\Big) \Gamma_{1}^{-4}\,E_{\textrm{kin},53}\,A_{*,-1}^{-1} \; \rm{days}, & k=2, 
\end{array} \right.
\end{align}
with $k$\,$=$\,$0$ for a uniform density $n$ environment, and $k$\,$=$\,$2$ and $A_*$ representing a wind-like environment \citep[e.g.,][]{ChevalierLi2000}. 
In Figure \ref{fig:alloweddecel} we show the allowed parameter space in terms of Lorentz factor and kinetic energy for both an ISM and wind environment. The allowed range of space, which depends also on density and (marginally) redshift, favors low Lorentz factor outflows with $\Gamma$\,$\lesssim$\,$10$ for either environment when considering reasonable densities. 

If the jet is not ultra relativistic, instead the deceleration time (for a uniform density environment) can be estimated as \citep{NakarPiran2011} 
\begin{align}
\label{eqn:tdecslow}
   t_\textrm{dec,s} \approx 600\,\Big(\frac{1+z}{2}\Big) E_\textrm{kin,52}^{1/3} n^{-1/3} \beta^{-5/3}\;\textrm{days}, 
\end{align} 
where $c\beta$ is the initial velocity of the outflow. For a reasonable range of energies and densities the deceleration in this regime can be significantly larger than $260$ d, especially if $\beta$\,$\ll$\,$1$ (see Figure \ref{fig:alloweddecel}). This is also the case for a wind environment. In either situation, the allowed range of $\beta$ is completely unconstrained below $\beta$\,$<$\,$0.9$ for kinetic energies as low as $10^{51}$ erg.

Therefore, there are multiple plausible interpretations for the lack of radio detections in our VLA and ATCA observations. Thus, we predict that luminous radio emission may become detectable from EP240408a on the timescale of a few hundred days (e.g., AT2018hyz), similar to other TDEs. The detection of late-time radio emission would exclude a GRB interpretation. Further late-time radio observations can aid in determining the nature of EP240408a.

\subsection{Interpretation of the Spectral Break}
\label{sec:specbreak}

Using the X-ray data obtained by \textit{NICER}, we are able to model the time-resolved X-ray spectra (\S \ref{sec:nicerspectra}) with an absorbed broken powerlaw and measure a time-varying spectral break between $3$\,$-$\,$6$ keV  (observer frame) as detailed in Figure \ref{fig:xrayspectra}. The exact time-evolution evolution of the spectral break is not well constrained due to declining data quality as the source fades, but it does not appear in the earliest \textit{NICER} data. As shown in Figure \ref{fig:xrayspectra} (bottom panels), the first appearance of the spectral break is in the third epoch (E3) of \textit{NICER} data at $\sim$\,3.4 d (observer frame). The spectral break energy appears relatively fixed in energy within errors, and does not display drastic changes between epochs. The median value is around $E_\textrm{break}$\,$\approx$\,$4.5$ keV (observer frame).

Spectral breaks are a common feature of synchrotron radiation \citep{Granot2002} and can be due to the passage of a few characteristic frequencies through the observed band. These breaks are commonly observed in GRBs, and also in some relativistic jetted TDEs \citep[e.g., AT2022cmc;][]{Yao2024}. Synchrotron radiation can arise from both internal and external processes to the jet. 
In an external shock origin, the characteristic synchrotron frequencies, such as the cooling frequency $\nu_\textrm{c}$ and the peak frequency $\nu_\textrm{m}$ have a well known temporal dependence \citep{Granot2002} that has also been measured in GRB afterglows. The time dependence of the cooling frequency is modified depending on the external environment with $\nu_\textrm{c}$\,$\propto$\,$t^{-1/2}$ for a uniform medium and $\nu_\textrm{c}$\,$\propto$\,$t^{0.5}$ for a wind-like medium, neglecting inverse Compton corrections \citep{Sari2001,Zou2009,Beniamini2015}. Instead, the peak frequency $\nu_\textrm{m}$ does not depend on the surrounding environment $\nu_\textrm{m}$\,$\propto$\,$t^{-3/2}$. However, a declining spectral break is disfavored by our analysis, especially one with such a strong time evolution. 

While the exact evolution of the spectral break observed in EP240408a is uncertain, it tends to be roughly consistent with a $t^{0.5}$ within errors (and we note the time dependence can be modified if the value of $k$ deviates from $k$\,$=$\,$2$). This is mainly due to the fact that the break energy is most well constrained at $\sim$\,$3.4$ d, and seems to increase following this based on the agreement between the values measured in the latter epochs (E5 to E8; Figure \ref{fig:xrayspectra}). In this case, it may well be due to the increase of the cooling frequency in a wind-like environment. However, it must be noted that the observed change in spectral index of the broken power-law above and below the spectral break is on the order of $0.5$\,$-$\,$1.0$, whereas the change in slope due to the passage of the cooling frequency should be $0.5$. In addition, the lightcurve does not follow the expected temporal behavior of an external shock synchrotron afterglow, especially due to the very steep decay at later times.

Instead, the observed X-ray emission may be due to non-thermal radiation from processes internal to the jet (such as internal shocks) at small radii (as required due to the extremely steep decay). If we assume the Lorentz factor of the emitting material is constant and adopt that the radius goes as $R$\,$\sim$\,$\Gamma^2 t$ we can estimate that $\nu_\textrm{c}$\,$\propto$\,$R^{(3/2)k-2}$\,$\propto$\,$t^{(3/2)k-2}$, which decreases for a uniform medium and increases for a wind-like medium, as in the external shocks scenario but with a steeper time dependence in both cases \citep{Granot2002}. As $k$ does not have to exactly satisfy either assumed density profile, the time evolution of $\nu_\textrm{c}$ can easily span a wide range and be consistent with our observations of the spectral break.  

We note that a spectral break between $10$\,$-$\,$15$ keV was observed in the relativistic TDE AT2022cmc \citep{Yao2024}. The temporal dependence there was roughly consistent with $\nu_\textrm{c}$\,$\propto$\,$t^{0.5}$ between 7.8 and 17.6 d (observer frame). This evolution could not be confirmed in the third epoch at 36.2 d, potentially due to the decreased data quality at this later time.

While the spectral break is most easily explained in a synchrotron emission model, and in general we favor a relativistic jetted TDE explanation for EP240408a, it is also possible to explain in the models proposed for the X-ray plateau in a GRB interpretation. For example, in the dissipative photosphere model (\S \ref{sec:photospheregrb}) a fixed spectral break easily matches the model prediction. As the geometric timescale (Equation \ref{eqn:tgeo}) is significantly shorter than the observed decay timescale, any variability, even of the spectral break, is directly impacted by the central engine activity and is therefore hard to completely exclude due to the wide range of possibilities for the engine's behavior (even from observations of GRBs). In the dissipative photosphere model, a time varying spectral break $E_\textrm{p}$\,$\propto$\,$(1+\sigma)^{1/2}\varepsilon_\textrm{rad}^{1/2}\Gamma^2$ could be potentially explained by a time evolution of the magnetization $\sigma$, radiation efficiency $\varepsilon_\textrm{rad}$, or outflow Lorentz factor $\Gamma$. However, envisioning a time evolution of these parameters that leads to an increasing break energy $E_\textrm{p}$ is less clear and may disfavor this interpretation for the plateau.

\subsection{Comparison of Intrinsic Rates}

We compare the intrinsic rate of long GRBs and jetted TDEs.
The inferred local rate of long GRBs is $79^{+57}_{-33}$ Gpc$^{-3}$ yr$^{-1}$ \citep{Ghirlanda2022} with a redshift evolution of $(1+z)^{3.2}$ out to $z$\,$\approx$\,$3$. Jetted TDE rates are more uncertain due to their smaller population and range between $\sim$\,$0.3$\,$-$\,$20$  Gpc$^{-3}$ yr$^{-1}$ for Sw J1644+57 and AT2018hyz, respectively \citep{Piran2023}. We  adopt the intrinsic rate of the jetted TDE AT2018hyz at $z$\,$=$\,$0.0457$ \citep{Cendes2022,Piran2023,Sfaradi2024} which is the closest and therefore leads to a higher (beaming corrected) rate of $\sim$\,$20$  Gpc$^{-3}$ yr$^{-1}$. This is consistent at the $2\sigma$ level with the rate of long GRBs at a similar redshift \citep{Ghirlanda2022}. The inferred rate of AT2018hyz would be significantly higher if interpreted as an on-axis jet \citep{Cendes2022}, not to mention the significant selection effects against detecting a similar event. This comparison implies that both jetted TDEs and long GRBs may occur at similar rates in the local Universe. However, EP240408a likely occurred at significantly higher redshifts $z$\,$\approx$\,$1$\,$-$\,$2$ in which case the intrinsic long GRB rate is higher by a factor of $\sim$\,$10$\,$-$\,$30$ whereas the rate of TDEs likely decreases rapidly \citep{Kochanek2016}.

\section{Conclusions}
\label{sec: conclusions}

In this work we have presented the results of our extensive multi-wavelength (X-ray, ultraviolet, optical, near-infrared, and radio) follow-up campaign of the \textit{Einstein Probe} transient EP240408a. Our campaign, which includes deep Gemini observations, uncovered a possible host galaxy association. However, the faint nature of the host leaves the distance scale and luminosity of EP240408a unconstrained. Based on the host brightness, we favor higher redshifts ($z$\,$\gtrsim$\,$0.5$) where the peak X-ray luminosity exceeds 10$^{49}$ erg s$^{-1}$. We note, however, that the lack of subarcsecond localization precludes both an accurate diagnostic of the host association or the possibility that the  candidate uncovered by Gemini is potentially an unrelated foreground galaxy.

The observed properties of EP240408a, such as the long-lived duration ($\sim$\,5 d), (likely) high X-ray luminosity, and lack of bright radio emission, do not directly align with any known transient class at any likely redshift. We have considered a variety of interpretations for the multi-wavelength dataset, and favor a peculiar GRB or jetted TDE at high-$z$ ($z$\,$\gtrsim$\,$1$), though neither perfectly explains the observations. The distinguishing factor between these two scenarios will be the detection or non-detection of radio emission on the timescale of hundreds of days. In addition, measuring the distance scale (redshift) of the candidate host galaxy, which may require space-based observations (e.g., \textit{Hubble Space Telescope} or \textit{James Webb Space Telescope}), is critical to the interpretation of EP240408a. Future follow-up of \textit{Einstein Probe} transients on rapid timescales may reveal more events falling into this rare, and potentially new, class of transient and aid in determining its true nature.

\acknowledgments

The authors thank the EP team, in particular Nanda Rea, Weimin Yuan, Wenda Zhang, and Chichuan Jin, for useful discussions. 
The authors acknowledge Jimmy DeLaunay regarding \textit{Swift}/BAT and \textit{Fermi}/GBM, Gaurav Waratkar regarding \textit{AstroSat}/CZTI, and Dmitry Svinkin regarding \textit{Konus-Wind}. 
B. O. acknowledges useful discussions with Phil Evans regarding the XRT data, Kathleen Labrie regarding the Gemini data, and Gourav Khullar and Anna O'Grady regarding the interpretation of the optical spectra. 
J. H. thanks George Younes for useful discussions on the nature of the source. 

B. O. is supported by the McWilliams Postdoctoral Fellowship at Carnegie Mellon University. J. H. acknowledges support from NASA under award number 80GSFC21M0002. P. B. is supported by a grant (no. 2020747) from the United States-Israel Binational Science Foundation (BSF), Jerusalem, Israel, by a grant (no. 1649/23) from the Israel Science Foundation and by a grant (no. 80NSSC 24K0770) from the NASA astrophysics theory program. M.N. is a Fonds de Recherche du Quebec – Nature et Technologies (FRQNT) postdoctoral fellow. S. S. is partially supported by LBNL Subcontract 7707915.  G. B. acknowledges funding from the European Union’s Horizon 2020 programme under the AHEAD2020 project (grant agreement no. 871158). Research at Perimeter Institute is supported in part by the Government of Canada through the Department of Innovation, Science and Economic Development and by the Province of Ontario through the Ministry of Colleges and Universities.

The National Radio Astronomy Observatory is a facility of the National Science Foundation operated under cooperative agreement by Associated Universities, Inc. 
Based on observations obtained at the international Gemini Observatory, a program of NSF's OIR Lab, which is managed by the Association of Universities for Research in Astronomy (AURA) under a cooperative agreement with the National Science Foundation on behalf of the Gemini Observatory partnership: the National Science Foundation (United States), National Research Council (Canada), Agencia Nacional de Investigaci\'{o}n y Desarrollo (Chile), Ministerio de Ciencia, Tecnolog\'{i}a e Innovaci\'{o}n (Argentina), Minist\'{e}rio da Ci\^{e}ncia, Tecnologia, Inova\c{c}\~{o}es e Comunica\c{c}\~{o}es (Brazil), and Korea Astronomy and Space Science Institute (Republic of Korea). 
This work is based on observations obtained with the Southern African Large Telescope. 

This research has made use of the NuSTAR Data Analysis Software (NuSTARDAS) jointly developed by the ASI Space Science Data Center (SSDC, Italy) and the California Institute of Technology (Caltech, USA). 
This work made use of data supplied by the UK \textit{Swift} Science Data Centre at the University of Leicester. 
This research has made use of the XRT Data Analysis Software (XRTDAS) developed under the responsibility of the ASI Science Data Center (ASDC), Italy. 
This research has made use of data and/or software provided by the High Energy Astrophysics Science Archive Research Center (HEASARC), which is a service of the Astrophysics Science Division at NASA/GSFC.
Some of the data presented herein were obtained at the W. M. Keck Observatory, which is operated as a scientific partnership among the California Institute of Technology, the University of California and the National Aeronautics and Space Administration. 

The Observatory was made possible by the generous financial support of the W. M. Keck Foundation. The authors wish to recognize and acknowledge the very significant cultural role and reverence that the summit of Maunakea has always had within the indigenous Hawaiian community. The Australia Telescope Compact Array is part of the Australia Telescope National Facility which is funded by the Australian Government for operation as a National Facility managed by CSIRO. We acknowledge the Gomeroi people as the Traditional Owners of the Observatory site.

%

\vspace{5mm}
\facilities{\textit{NuSTAR}, \textit{NICER}, \textit{Swift}, Gemini, Keck, DECam, VLA, ATCA}


\software{\texttt{HEASoft}, \texttt{XRTDAS}, \texttt{NuSTARDAS}, \texttt{NICERDAS}, \texttt{XSPEC} \citep{Arnaud1996}, \texttt{Dragons} \citep{Labrie2019,Labrie2023}, \texttt{SFFT} \citep{Hu2022}, \texttt{IRAF} \citep{IRAF1986}, \texttt{CASA} \citep{CASA2007}, \texttt{miriad} \citep{Sault1995_MIRIAD}, \texttt{pysynphot} \citep{pysynphot}, \texttt{Astropy} \citep{Astropy2018} }





\bibliography{biblio}{}
\bibliographystyle{aasjournal}



\appendix
\setcounter{table}{0}
\renewcommand{\thetable}{A\arabic{table}}
\setcounter{figure}{0}
\renewcommand{\thefigure}{A\arabic{figure}}

\section{Log of Observations}

Here, we present the log of X-ray, ultraviolet,  optical, infrared, and radio observations analyzed in this work. 

\begin{table}[ht]
\centering
\caption{Log of X-ray observations used in this work. 
}
\label{tab: observationsXray}
\begin{tabular}{lcccccc}
\hline\hline
\textbf{Start Time (UT)} & \textbf{$\delta T$ (d)} & \textbf{Telescope} & \textbf{Instrument} &  \textbf{Exposure (s)} & \textbf{ObsID}  \\
\hline
2024-04-10 13:33:08.00 & 1.82 & \textit{NICER} & XTI & 2216 & 7204340101 \\ 
2024-04-11 00:23:30.00 & 2.27 & \textit{NICER} & XTI & 6064 & 7204340102 \\ 
2024-04-12 01:10:19.00 & 3.30 & \textit{NICER} & XTI & 8807 & 7204340103 \\ 
2024-04-13 00:25:23.00 & 4.27 & \textit{NICER} & XTI & 5277 & 7204340104 \\ 
2024-04-14 01:10:54.00 & 5.30 & \textit{NICER} & XTI & 4856 & 7204340105 \\ 
2024-04-15 00:24:35.00 & 6.30 & \textit{NICER} & XTI & 3488 & 7204340106 \\ 
2024-04-16 16:40:37.00 & 7.94 & \textit{NICER} & XTI & 331 & 7204340107 \\
2024-04-17 01:57:57.00 & 8.33 & \textit{NICER} & XTI & 831 & 7204340108 \\
2024-04-18 01:11:16.00 & 9.30 & \textit{NICER} & XTI & 2455 & 7204340109 \\ 
2024-04-19 01:57:39.00 & 10.33 & \textit{NICER} & XTI & 1560 & 7204340110 \\ 
2024-04-20 01:08:00.00 & 11.30 & \textit{NICER} & XTI & 4405 & 7204340111 \\ 
2024-04-21 00:22:56.00 & 12.27 & \textit{NICER} & XTI & 4398 & 7204340112 \\ 
2024-04-22 01:10:20.00 & 13.30 & \textit{NICER} & XTI & 4171 & 7204340113 \\ 
2024-04-23 00:23:00.00 & 14.27  & \textit{NICER} & XTI & 2700 & 7204340114 \\ 
2024-04-24 01:08:00.00 & 15.30 & \textit{NICER} & XTI & 2334 & 7204340115 \\ 
2024-04-25 00:21:39.00 & 16.27 & \textit{NICER} & XTI & 3136 & 7204340116 \\ 
2024-04-26 04:12:59.00 &  17.43 & \textit{NICER} & XTI & 6647 & 7204340117 \\ 
2024-04-27 00:20:01.00 & 18.27 & \textit{NICER} & XTI & 8229 & 7204340118 \\ 
2024-04-28 01:04:49.00 & 19.29 & \textit{NICER} & XTI & 5443 & 7204340119 \\ 
2024-04-29 00:17:30.00 & 20.27 & \textit{NICER} & XTI & 6846 & 7204340120 \\ 
2024-04-30 01:00:00.00 & 21.29 & \textit{NICER} & XTI & 4611 & 7204340121 \\ 
2024-05-01 00:12:40.00 & 22.26 & \textit{NICER} & XTI & 6797 & 7204340122 \\ 
2024-05-02 01:02:00.00 & 23.30 & \textit{NICER} & XTI & 4070 & 7204340123 \\ 
2024-05-03 00:15:40.00 & 24.26 & \textit{NICER} & XTI & 2939 & 7204340124 \\ 
2024-05-09 03:38:01.00 & 30.40 & \textit{NICER} & XTI & 251 & 7204340125 \\
2024-05-10 22:32:07.00 & 32.19 & \textit{NICER} & XTI & 468 & 7204340126 \\
2024-05-11 00:05:07.00 & 32.26 & \textit{NICER} & XTI & 1378 & 7204340127 \\
2024-05-12 00:50:29.00 &  33.29 & \textit{NICER} & XTI & 1312 & 7204340128 \\
2024-05-14 11:39:20.00 & 35.74 & \textit{NICER} & XTI & 1151 & 7204340129 \\
2024-05-15 09:19:19.00 & 36.64 & \textit{NICER} & XTI & 948 & 7204340130 \\
2024-05-16 10:05:40.00 & 37.67 & \textit{NICER} & XTI & 883 & 7204340131 \\	
 \hline
 2024-04-10 02:54:18 & 1.40 & \textit{Swift} & XRT  & 1825 & 16599001 \\
 2024-04-19 03:38:43 & 10.40 & \textit{Swift} & XRT  & 1015 &
 16599003 \\
 2024-04-21 03:00:57	 & 12.38 & \textit{Swift} & XRT  & 2648 &
 16599004 \\
  2024-04-24 03:50:57 & 15.41 & \textit{Swift} & XRT  & 965&
 16599005 \\
  2024-04-26 04:37:57 & 17.45 & \textit{Swift} & XRT  & 2451 &
 16599006 \\
 \hline
 2024-04-22 00:36:09 & 13.28 & \textit{NuSTAR} & FPMA/B & 42553 & 91001622002 \\
\hline\hline
\end{tabular}
\end{table}

\begin{table}[ht]
\centering
\caption{Log of optical and near-infrared observations used in this work. Photometry is reported in the AB magnitude system and is not corrected for Galactic extinction, which is tabulated for each filter in the $A_\lambda$ column corresponding to the line-of-sight Galactic extinction of $E(B-V)$\,$=$\,$0.076$ mag \citep{Schlafly2011}. Upper limits are reported at $3\sigma$.
}
\label{tab: observationsOpt}
\begin{tabular}{lccccccccc}
\hline\hline
\textbf{Start Time} & \textbf{$\delta T$} & \textbf{Telescope} & \textbf{Instrument} &  \textbf{Exposure} & \textbf{Filter} & \textbf{Transient} & \textbf{Candidate Host} & \textbf{$A_\lambda$} \\
\textbf{(UT)} & \textbf{(d)} &  &  &  \textbf{(s)} & & \textbf{(mag)} & \textbf{(mag)} & \textbf{(mag)} \\
\hline
\hline
    \multicolumn{9}{c}{\textbf{Gemini}}  \\
\hline
2024-04-25 03:34:24 & 16.40 & Gemini & GMOS-S & 720 &  \textit{r} & $>$25.1$^a$ & -- & 0.21 \\
2024-04-25 03:57:56 & 16.42 & Gemini & GMOS-S & 720 & \textit{i} & $>$25.0$^a$ & -- & 0.16 \\
2024-04-25 04:31:40 & 16.44 & Gemini & F2 & 900 & \textit{J} & $>$23.1$^a$ & -- & 0.07 \\
2024-05-02 23:05:38  & 24.21 & Gemini & GMOS-S & 720 &  \textit{r} & $>$25.5$^a$ & -- &  0.21 \\
2024-05-02 23:25:34 & 24.23 & Gemini & GMOS-S & 720 & \textit{i} & $>$25.6$^a$ & -- & 0.16 \\
2024-05-03 23:02:58 & 25.21 & Gemini & F2 & 900 & \textit{J} & $>$23.2 & $>$23.2 & 0.07 \\
2024-06-30 00:29:04 & 82.27 & Gemini & GMOS-S & 960 & \textit{r} & $>$26.0$^b$ & $24.2\pm0.1$ & 0.21 \\
2024-06-30 00:59:36 & 82.29 & Gemini & GMOS-S & 1000 & \textit{i} & $>$26.1$^b$ & $23.9\pm0.1$ & 0.16 \\
\hline \hline
    \multicolumn{9}{c}{\textbf{UVOT}}  \\
\hline
2024-04-10 03:08:42 & 1.38 & \textit{Swift} & UVOT & 157 & \textit{v} & $>$19.23 & $>$19.23 & 0.25 \\
2024-04-10 03:01:52 & 1.38 & \textit{Swift} & UVOT & 157 & \textit{b} & $>$20.16 & $>$20.16 & 0.32 \\
2024-04-10 03:00:28 & 1.38 & \textit{Swift} & UVOT & 157 & \textit{u} & $>$ 20.96 & $>$20.96 & 0.40 \\
2024-04-10 02:57:44 & 1.38 & \textit{Swift} & UVOT & 314 & \textit{uvw1} & $>$20.99 & $>$20.99 & 0.56 \\
2024-04-10 03:10:07 & 1.38 & \textit{Swift} & UVOT & 331 & \textit{uvm2} & $>$21.54 & $>$21.54 & 0.71 \\
2024-04-10 03:03:18 & 1.38 & \textit{Swift} & UVOT & 629 & \textit{uvw2} & $>$22.18 & $>$22.18 & 0.64 \\
2024-04-20 03:26:05 & 11.4 & \textit{Swift} & UVOT & 108 & \textit{u} & $>$20.73 & $>$20.73 & 0.40 \\
2024-04-19 03:41:31 & 10.41 & \textit{Swift} & UVOT & 884 & \textit{uvw1} & $>$20.02 & $>$20.02 & 0.56  \\
2024-04-21 03:00:57 & 12.38 & \textit{Swift} & UVOT & 2598 & \textit{uvw2} & $>$22.94 & $>$22.94 & 0.64 \\
2024-04-24 03:50:57 & 15.41 & \textit{Swift} & UVOT & 994 & \textit{u} & $>$22.11 & $>$22.11 & 0.40 \\
2024-04-26 04:37:57 & 18.42 & \textit{Swift} & UVOT & 2181 & \textit{uvw1} & $>$22.42 & $>$22.42 & 0.56 \\
2024-04-27 04:08:18 & 18.42 & \textit{Swift} & UVOT & 228 & \textit{uvm2} & $>$21.23 & $>$21.23 & 0.71 \\
\hline \hline
    \multicolumn{9}{c}{\textbf{Stacked UVOT}}  \\
\hline
2024-04-10 02:54:18 & 1.4-12.4 & \textit{Swift} & UVOT & 3227 &  \textit{uvw2} & $>$23.1 & $>$23.1 & 0.64 \\
2024-04-10 02:54:18 & 1.4-18.4 & \textit{Swift} & UVOT & 559 & \textit{uvm2} & $>$21.9 & $>$21.9 & 0.71 \\
2024-04-10 02:54:18 & 1.4-18.4 & \textit{Swift} & UVOT & 3379 & \textit{uvw1} & $>$22.7 & $>$22.7 & 0.56 \\
2024-04-10 02:54:18 & 1.4-15.4 & \textit{Swift} & UVOT & 1259 & \textit{u} & $>$22.2 & $>$22.2 & 0.40 \\
\hline \hline
\end{tabular}
\tablecomments{ $^a$ Upper limits from \texttt{SFFT} image subtraction. $^b$ Upper limits from a source free region within the XRT localization.}
\end{table}

\begin{table}[ht]
\centering
\caption{Log of radio observations used in this work. Upper limits are reported at the $3\sigma$ level.  
}
\label{tab: observationsradio}
\begin{tabular}{lccccccc}
\hline\hline
   \textbf{Start Date (UT)}  & \textbf{$\delta T$ (d)} & \textbf{Telescope} & \textbf{Configuration}  & \textbf{Band} & \textbf{Exposure (s)} &  \textbf{Flux Density ($\mu$Jy)}\\
\hline
2024-04-19 02:08:22 & 10.34 & VLA & C & X (10 GHz)   & 1410 & $<17$ \\
2024-05-01 04:51:05 & 22.45 & ATCA & -- & C (5.5 GHz) & 5760 & $<60$ \\
2024-05-01 04:51:05 & 22.45 & ATCA & -- & X (9 GHz) & 5760 & $<60$ \\
2024-05-10 06:39:45 & 31.53 & ATCA & -- & K (18 GHz) & 5400 & $<180^a$ \\
2024-09-13 17:40:28 & 158.0 & VLA & B & X (10 GHz)   & 1410 & $<21$ \\
2024-12-12 10:29:00 & 247.7 & VLA & A & X (10 GHz)   & 1410 & $<20$ \\
\hline\hline
\end{tabular}
\tablecomments{ $^a$ This is a rough estimate of the upper limit at this epoch in a merged observation, see \S \ref{sec: ATCA} for caveats.}
\end{table}

\section{Timing Analysis/Pulsations search on \nicer Data}
\label{sec:timing}
We conducted an X-ray pulsation search in the event that a neutron star is responsible for the transient event, such as a millisecond pulsar or a magnetar. In particular, we utilized the acceleration search, a Fourier domain technique that accounts for the ``smearing'' of a potential coherent signal due to the orbital Doppler modulation \citep{Ransom2002}. In Fourier space, the signal would be smeared across $z$\,$=$\,$\alpha hfT^2/c$ Fourier bins, where $\alpha$ is the pulsar acceleration, $h$ is the harmonic number ($h$\,$=$\,$1$ is the fundamental), $f$ is the potential pulse frequency, $T$ is the length of the time segment being searched over, and $c$ is the speed of light \citep{Ransom2002}. Acceleration searches are optimal when the pulsar acceleration is roughly constant within time segments such that $T \leq P_{\rm orb}/10$, where $P_{\rm orb}/10$ is the orbital period of the binary system. We note that for isolated pulsars (e.g., a magnetar), $z=0$. We employed the \texttt{accelsearch} routine as implemented within version 4.0 of \texttt{PRESTO}\footnote{\url{https://github.com/scottransom/presto}} for the acceleration search \citep{Ransom2002,PRESTO2011}, searching over 0.5--1000~Hz and between 0.5--2.0~keV and 2.0--10.0~keV. We also searched up to a maximum of 100 bins in Fourier frequency space that the Doppler modulated signal would drift across. We did not find any significant periodicity candidates from the acceleration search. 

We also constructed an averaged power spectrum using the \texttt{AveragedPowerspectrum} class in \texttt{Stingray} \citep{Stingray2019_ApJ,Stingray2019_JOSS,Stingray2024_Zenodo}, where we used 256~s bins and a bin size of $\Delta t = 2^{-12}{\rm\,s}$. We did not find any significant coherent periodicity in the averaged spectrum, and we determined a $3\sigma$ upper limit on the sinusoidal pulsation amplitude of 24.5--27.0$\%$ over 1--1000~Hz \citep{Vaughan1994}.

\section{Notes on Candidate Optical Counterparts}
\label{sec:counter}

\begin{figure*}
    \centering
\includegraphics[width=0.45\columnwidth]{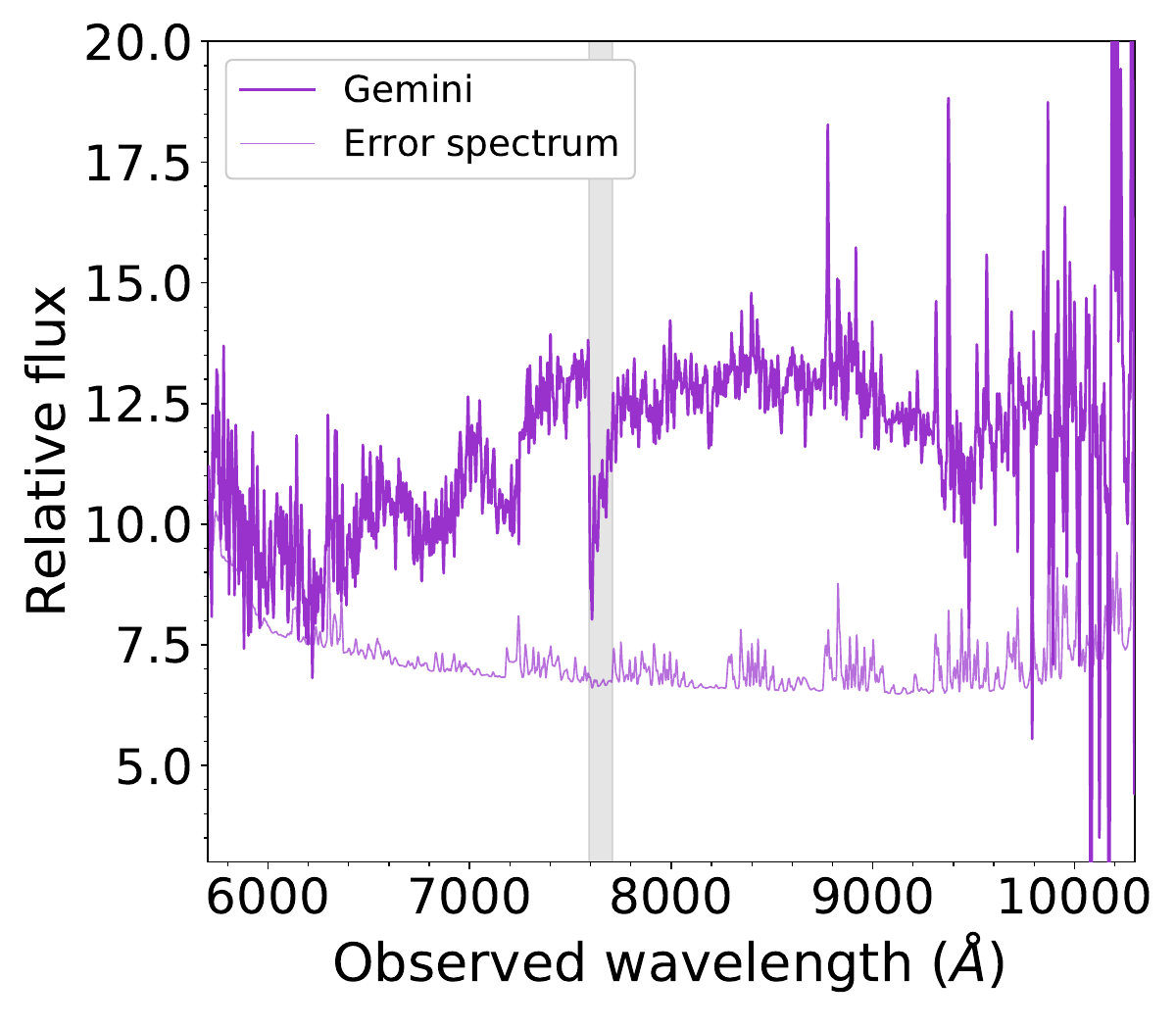}
\includegraphics[width=0.45\columnwidth]{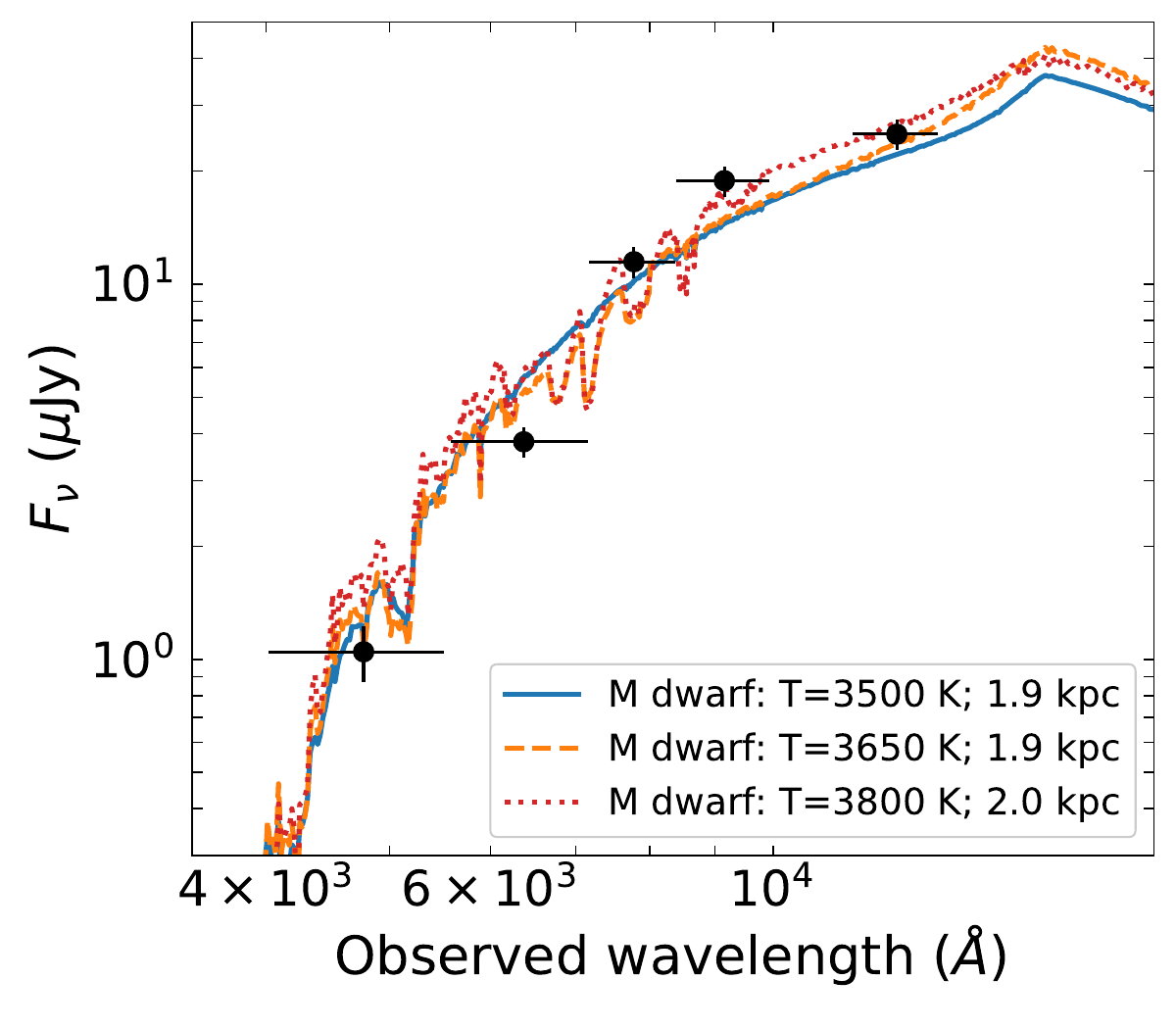}
    \caption{\textbf{Left:} Gemini GMOS-S optical spectrum of the bright source \citep{GCN36059grond}  lying within the initial (standard) XRT localization. The error spectrum is also shown. The spectrum is smoothed with a Savitzky-Golay filter of 3 pixels for display purposes. Vertical gray shaded regions mark atmospheric telluric absorption regions. 
    The sharp peaks in the observed spectrum above $8,600$\,\AA\, are due to sky emission lines and are not real features. \textbf{Right:} Spectral energy distribution of this source \citep{GCN36059grond} compared to Kurucz stellar models \citep{Kurucz1993}. We find adequate matches to the observed photometry corresponding to late spectral type M dwarfs at $\sim$ 2 kpc. The models are extincted by $E(B-V)$\,$=$\,$0.076$ mag \citep{Schlafly2011}.} 
    \label{fig:spectra}
\end{figure*}

The bright source reported by \citet{GCN36059grond} appears point-like in our images (Figure \ref{fig:fc}), indistinguishable from the numerous field stars in our deep Gemini images. While this source lies in the initial XRT position, it is no longer consistent with the updated enhanced position (Figure \ref{fig:fc}). In any case, the broad absorption features observed in our Gemini spectrum (Figure \ref{fig:spectra}; left panel) between $6,000$\,$-$\,$7,000$ \AA\, are indicative of a stellar spectrum. \citet{ATel16589saltnicer} carried out follow-up spectroscopy with SALT and reported a potential emission line from this source. However, a reanalysis of the spectrum shows that this is unrelated to this source. 

We compared the ultraviolet, optical, and near-infrared  spectral energy distribution (SED) to stellar models \citep{Kurucz1993} using \texttt{pysynphot} \citep{pysynphot}. We find an appropriate match to the SED for a late spectral type dwarf of type between M2V and M4V located at a distance of $\sim$2 kpc. We consider an M dwarf with temperature 3500 K and metallicity $\log(Z/Z_\odot)$\,$=$\,$-2$ (solid line), with temperature 3600 K with  $\log(Z/Z_\odot)$\,$=$\,$-0.5$ (dashed line), and with temperature 3800 K with  $\log(Z/Z_\odot)$\,$=$\,$0.5$ (dotted line). This SED comparison is shown in Figure \ref{fig:spectra} (right panel). This conclusively confirms the source is unrelated to EP240408a.

\end{document}